\documentclass[aps,superscriptaddress,pra,twocolumn,,showpacs]{revtex4-1}
\usepackage{amsfonts, amsmath, amssymb, bm, bbm, graphicx, epsfig, epstopdf}
\usepackage{braket}
\usepackage{dcolumn}
\usepackage{textcomp}
\usepackage{hyperref}
\usepackage{array}
\usepackage{color}
\definecolor{forestgreen}{RGB}{34,139,34}

\usepackage{lipsum}

\usepackage[normalem]{ulem} 
\usepackage{color} 

\begin{document}
	\title{A first-principles calculation of electron-phonon interactions for the $\text{C}_2\text{C}_\text{N}$ and $\text{V}_\text{N}\text{N}_\text{B}$ defects in hexagonal boron nitride}
	\author{K. Sharman}
	\affiliation{Institute for Quantum Science and Technology, 2500 University Drive NW, Calgary, Alberta T2N 1N4, Canada}
	\affiliation{Department of Physics and Astronomy, University of Calgary, 2500 University Drive NW, Calgary, Alberta T2N 1N4, Canada}
	\affiliation{Hotchkiss Brain Institute, University of Calgary, 3330 Hospital Drive NW, Calgary, Alberta T2N 4N1, Canada}
	\author{O. Golami}
	\affiliation{Institute for Quantum Science and Technology, 2500 University Drive NW, Calgary, Alberta T2N 1N4, Canada}
	\affiliation{Department of Physics and Astronomy, University of Calgary, 2500 University Drive NW, Calgary, Alberta T2N 1N4, Canada}
	\affiliation{Hotchkiss Brain Institute, University of Calgary, 3330 Hospital Drive NW, Calgary, Alberta T2N 4N1, Canada}
	\author{S. C. Wein}
	\affiliation{Institute for Quantum Science and Technology, 2500 University Drive NW, Calgary, Alberta T2N 1N4, Canada}
	\affiliation{Department of Physics and Astronomy, University of Calgary, 2500 University Drive NW, Calgary, Alberta T2N 1N4, Canada}
	\affiliation{Hotchkiss Brain Institute, University of Calgary, 3330 Hospital Drive NW, Calgary, Alberta T2N 4N1, Canada}
	\affiliation{Quandela SAS, 10 Boulevard Thomas Gobert, 91120, Palaiseau, France}
	\author{H. Zadeh-Haghighi}
	\affiliation{Institute for Quantum Science and Technology, 2500 University Drive NW, Calgary, Alberta T2N 1N4, Canada}
	\affiliation{Department of Physics and Astronomy, University of Calgary, 2500 University Drive NW, Calgary, Alberta T2N 1N4, Canada}
	\affiliation{Hotchkiss Brain Institute, University of Calgary, 3330 Hospital Drive NW, Calgary, Alberta T2N 4N1, Canada}
	\author{C. G. Rocha}
	\affiliation{Institute for Quantum Science and Technology, 2500 University Drive NW, Calgary, Alberta T2N 1N4, Canada}
	\affiliation{Department of Physics and Astronomy, University of Calgary, 2500 University Drive NW, Calgary, Alberta T2N 1N4, Canada}
	\affiliation{Hotchkiss Brain Institute, University of Calgary, 3330 Hospital Drive NW, Calgary, Alberta T2N 4N1, Canada}
	\author{A. Kubanek}
	\affiliation{Institute for Quantum Optics, Ulm University, D-89081 Ulm, Germany}
	\affiliation{Center for Integrated Quantum Science and Technology (IQst), Ulm University, D-89081 Ulm, Germany}
	\author{C. Simon}
	\affiliation{Institute for Quantum Science and Technology, 2500 University Drive NW, Calgary, Alberta T2N 1N4, Canada}
	\affiliation{Department of Physics and Astronomy, University of Calgary, 2500 University Drive NW, Calgary, Alberta T2N 1N4, Canada}
	\affiliation{Hotchkiss Brain Institute, University of Calgary, 3330 Hospital Drive NW, Calgary, Alberta T2N 4N1, Canada}
	\begin{abstract}
        Quantum emitters in two-dimensional hexagonal boron nitride (h-BN) have generated significant interest due to observations of ultra-bright emission made at room temperature. The expectation that solid-state emitters exhibit broad zero-phonon lines at elevated temperatures has been put in question by recent observations of Fourier transform (FT) limited photons emitted from h-BN flakes at room temperature. The mechanism responsible for the narrow lines has been suggested to be a mechanical decoupling from in-plane phonons due to an out-of-plane distortion of the emitter's orbitals. All decoupled emitters produce photons that are directed in-plane, suggesting that the dipoles are oriented perpendicular to the h-BN plane. Motivated by the promise of an efficient and scalable source of indistinguishable photons that can operate at room temperature, we have developed an approach using density functional theory (DFT) to determine the electron-phonon coupling for defects that have in- and out-of-plane transition dipole moments. Our DFT calculations reveal that the $\text{C}_2 \text{C}_\text{N}$ defect has an in-plane transition dipole moment, and that of the $\text{V}_\text{N} \text{N}_\text{B}$ defect is perpendicular to the plane. We exploit the two-dimensional framework recently implemented in \texttt{QUANTUM ESPRESSO} to determine both the phonon density of states and the electron-phonon matrix elements associated with the h-BN defective structures. We find no indication that an out-of-plane transition dipole is sufficient to obtain FT-limited photons at room temperature. Our work also provides direction to future DFT software developments and adds to the growing list of calculations relevant to researchers in the field of solid-state quantum information processing.
	\end{abstract}

\maketitle
\section{Introduction}
\label{Intro}

Single-photon sources (SPSs) are essential components of many quantum technologies, including linear optical quantum computing \cite{knill2001scheme,kok2007linear}, quantum random number generation \cite{jennewein2000fast, hoese2022single}, and quantum communication \cite{gisin2007quantum, sangouard2011quantum, wu2020near}. Atom-like solid-state emitters are regarded as promising SPSs because they combine the excellent optical properties of atoms with the reproducible and scalable fabrication methods that are available to solid-state materials \cite{somaschi2016near, loredo2016scalable, gale2021deterministic, koehl2015designing}. Among solid-state SPSs, emitters in two-dimensional (2D) hexagonal boron nitride (h-BN) \cite{he2015single, toth2019single, mendelson2021identifying} have recently generated significant interest due to experimental observations of ultra-bright emission at room temperature \cite{tran2016quantum, li2019near, kianinia2017robust, proscia2018near, jungwirth2016temperature, bourrellier2016bright, grosso2017tunable, zeng2022integrated}. 

Many quantum technologies which incorporate SPSs make use of single- or two-photon interference where indistinguishable photons are required. A significant disadvantage associated with solid-state systems such as h-BN is that phonon-induced dephasing typically results in homogeneous spectral broadening of the emission lines, which in turn, reduces the degree of photon indistinguishability \cite{white2021phonon}.

When the excited electronic state of an optical transition is subject to negligible broadening mechanisms, the zero-phonon line (ZPL) is said to be Fourier transform (FT) limited, meaning the spectral width is given by the radiative decay rate \cite{hogele2004voltage, houel2012probing, kuhlmann2013charge, kuhlmann2015transform}. SPSs which produce FT-limited photons are desired because they have the potential to serve as efficient sources of indistinguishable photons. Unfortunately, solid-state SPSs are typically required to operate at cryogenic temperatures to produce FT-limited emission \cite{tamarat2006stark, batalov2008temporal, kuhlmann2015transform}. The development of solid-state indistinguishable SPSs which can operate at room temperature would represent a major advancement as it would drastically reduce the difficulties associated with cryogenic cooling.

Recently, Dietrich \textit{et al.} \cite{dietrich2020solid} observed FT-limited photons from emitters in h-BN at room temperature. While the atomic structure of the emitter (or emitters) responsible is not yet known, it has been suggested that the underlying mechanism responsible for the narrow lines is an out-of-plane distortion of the emitter's orbitals and/or the transition dipole moment \cite{hoese2020mechanical}. Support for this claim is given by analysis of the measured photoluminescence spectrum which indicates that the emitter weakly couples to in-plane phonon modes. Additionally, the ability to observe emission from a mechanically decoupled emitter requires that the emitter is on a tilted plane, suggesting that the transition dipole is perpendicular to the plane.

Motivated by the desire to develop solid-state SPSs which are capable of efficiently providing indistinguishable photons at room temperature, we have developed an approach using first-principle calculations to gain insight into the temperature-dependent electron-phonon interactions that lead to a broadening of the ZPL emission. Density functional theory (DFT) calculations have been extensively used to predict the electronic, vibrational, and optical properties of numerous solid-state emitters \cite{hepp2014electronic, gali2008ab, abdi2018color, ping2021computational, jara2021first, bhang2021first, sajid2018defect, ivady2020ab}. A computational methodology for determining the electron-phonon coupling associated with an optical transition has been developed by Alkauskas \textit{et al.} \cite{alkauskas2014first, razinkovas2021vibrational}. The method was applied to the negatively charged nitrogen vacancy (NV$^-$) center in diamond, resulting in calculated luminescence and absorption lineshapes that are in excellent agreement with experimental data. Unfortunately, the model does not predict the homogeneous broadening of the ZPL, and the width is selected to match experimental observations. The temperature dependence of the ZPL width for the NV$^-$ center has, however, been successfully described using a model that considers interactions between the excited electronic states and acoustic phonons \cite{plakhotnik2015electron}. The key quantities required to calculate the linewidth include the phonon density of states (DOS) and the electron-phonon matrix elements. The analysis presented in Ref. \cite{plakhotnik2015electron} was based on a combination of theory and experiment. We perform a related investigation for defects in 2D h-BN without the need for experimental data, which is not yet available for the defects analyzed in this work.

To investigate the relationship between the transition dipole direction and the electron-phonon interaction strengths, we have studied two defects in 2D h-BN. Our DFT calculations reveal that the $\text{C}_2 \text{C}_\text{N}$ and $\text{V}_\text{N} \text{N}_\text{B}$ defects have transition dipoles that are in- and out-of-plane, respectively. We use density functional perturbation theory (DFPT), the goal of which is to compute derivatives of the DFT electronic energy with respect to different perturbations, to calculate and compare both the phonon DOS and the electron-phonon matrix elements of these two defects. Our findings provide no indication that an out-of-plane transition dipole moment will result in the low electron-phonon coupling required to produce FT-limited lines at room temperature.

The paper is organized as follows. In Sec. \ref{quantum_theory}, we discuss the electron-phonon interactions which are responsible for spectral broadening. In Sec. \ref{computational_details}, we provide the details of our calculations and present the results. In Sec. \ref{discussion}, we discuss the limitations associated with our analysis, along with possible extensions to our investigation. We conclude and provide an outlook in Sec. \ref{conclusion}.

\section{Theory}
\label{quantum_theory}

\subsection{Electron-phonon interactions}
\label{theory_electron_phonon}

In Kohn-Sham DFT formalism \cite{kohn1965self}, an interacting many-particle system is described in terms of an effective non-interacting system. The standard Hamiltonian which describes the electron-phonon coupling within DFT is obtained by expanding the Kohn-Sham effective potential in terms of nuclear displacements from their equilibrium positions, and is given by \cite{giustino2017electron}
\begin{equation}
\begin{split}
    \hat{H}_{\mathrm{ep}} &= \frac{1}{\sqrt{N_\mathrm{p}}} \sum_{\substack{\textbf{k}, \textbf{q} \\ m n \nu }} g_{mn\nu}^{(1)} (\textbf{k}, \textbf{q}) \hat{c}_{m, \textbf{k} + \textbf{q}}^\dagger \hat{c}_{n, \textbf{k}} (\hat{a}_{\nu, \textbf{q}} + \hat{a}_{\nu, -\textbf{q}}^\dagger) \\
    &+ \Bigg[ \frac{1}{N_\mathrm{p}} \sum_{\substack{\textbf{k}, \textbf{q}, \textbf{q}^\prime \\ m n \nu \nu^\prime }} g_{mn \nu \nu^\prime}^{(2)} (\textbf{k}, \textbf{q}, \textbf{q}^\prime) \hat{c}_{m, \textbf{k} + \textbf{q} + \textbf{q}^\prime}^\dagger \hat{c}_{n, \textbf{k}} \\
    &\times (\hat{a}_{\nu, \textbf{q}} + \hat{a}_{\nu, -\textbf{q}}^\dagger )(\hat{a}_{\nu^\prime, \textbf{q}^\prime} + \hat{a}_{\nu^\prime, -\textbf{q}^\prime}^\dagger )\Bigg].
\end{split}
\label{Hamiltonian}
\end{equation}
The first line of Eqn. \ref{Hamiltonian} describes, to first order in nuclear displacements, the linear coupling between an electron in band $n$ ($m$) with wave vector $\textbf{k}$ (\textbf{k}+\textbf{q}), and a phonon in branch $\nu$ with wave vector $\textbf{q}$ ($-\textbf{q}$). The electronic (phononic) creation and annihilation operators are given by $\hat{c}_{n, \textbf{k}}^\dagger$ and $\hat{c}_{n, \textbf{k}}$ ($\hat{a}_{\nu, \textbf{q}}^\dagger$ and $\hat{a}_{\nu, \textbf{q}}$), respectively. The phonon wave vectors define a uniform grid of $N_\mathrm{p}$ points in one unit cell of the reciprocal lattice. The second and third lines of Eqn. \ref{Hamiltonian} describe the additional quadratic electron-phonon coupling that arises when expansion of the Kohn-Sham potential includes second-order nuclear displacements. The quadratic terms describe the interactions between an electron and two phonons, where one phonon is in mode $\nu$ with wave vector $\textbf{q}$ and the other is in mode $\nu^\prime$ with wave vector $\textbf{q}^\prime$. The matrix elements $g_{mn\nu}^{(1)} (\textbf{k}, \textbf{q}) $ and $g_{mn \nu \nu^\prime}^{(2)} (\textbf{k}, \textbf{q}, \textbf{q}^\prime)$ quantify the magnitude of the linear and quadratic electron-phonon couplings, respectively.
 

Previous investigations \cite{maradudin1966solid, davies1974vibronic, fu2009observation, abtew2011dynamic, plakhotnik2015electron} into the temperature-dependent broadening of ZPL emission from solid-state defects have identified that two-phonon processes dominate at higher temperatures, but the specific transitions involved vary from defect to defect. In the negatively charged silicon-vacancy center in diamond, for example, linear interactions with phonons result in second-order elastic scattering processes that dominate the ZPL broadening at temperatures above 70 K \cite{jahnke2015electron}. In the NV$^-$ center, on the other hand, linear interactions with phonons result in second-order inelastic Raman-type scattering processes, and quadratic interactions with phonons result in first-order elastic scattering mechanisms \cite{plakhotnik2015electron}. Both types of processes must be considered when determining the ZPL width at room temperature.

While the exact nature of the electron-phonon interactions depends on the electronic structure of the defect, all these processes share the fact that the associated interaction rates, which are calculated using Fermi's Golden Rule \cite{sakurai2021modern}, are functions of both the phonon DOS and the electron-phonon matrix elements. Thus, calculating the electron-phonon matrix elements and the phonon DOS provides valuable insight into the processes responsible for the ZPL broadening.

The FT-limited emission reported in Ref. \cite{dietrich2020solid} is the result of either very small matrix elements or vanishing phonon DOS, or both. If defects with out-of-plane transition dipole moments do indeed result in FT-limited emission at elevated temperatures, then the corresponding electron-phonon coupling should be noticeably smaller than that of a defect center which possess an in-plane transition dipole.

To simplify the analysis, we do not consider electron-phonon interactions which result in transitions between electronic bands. We are interested in a system that displays no ZPL broadening when the temperature increases, requiring that both intraband and interband scattering processes are negligible. Additionally, phonon-assisted transitions between excited states are expected to be negligible for both defects, since the excited states are non-degenerate and well-separated in energy \cite{abdi2018color, golami2022b}. A good starting-point is therefore to determine the impact that the transition dipole direction has on only the intraband scattering rates.

\subsection{Phonon density of states}
\label{theory_phonon_dos}

In the Born-Oppenheimer approximation, the nuclei move in a potential energy ($U$) given by the total energy of the electronic system calculated at fixed nuclei. Dynamical properties of solid-state structures can be derived by expanding $U$ with respect to small displacements, $u_{\kappa \alpha}^l$, about the equilibrium configuration of the nuclei, where $\alpha$ are the cartesian indices of nucleus $\kappa$ in unit cell $l$. In the harmonic approximation, the expansion is given by
\begin{equation}
U  = U_0 + \frac{1}{2} \sum_{l \kappa \alpha} \sum_{l^\prime \kappa^\prime \alpha^\prime} \frac{\partial^2 U}{\partial u_{\kappa \alpha}^l \partial u_{\kappa^\prime \alpha^\prime}^{l^\prime} } u_{\kappa \alpha}^l u_{\kappa^\prime \alpha^\prime}^{l^\prime} .
\end{equation}
Here, $U_0$ is the total energy when the nuclei are in their equilibrium positions, and $\partial^2 U / \partial u_{\kappa \alpha}^l \partial u_{\kappa^\prime \alpha^\prime}^{l^\prime} = C_{\kappa \alpha,  \kappa^\prime \alpha^\prime}^{l, l^\prime} $ is the matrix of interatomic force constants (IFCs) \cite{heid201312, verstraete2014c}. 
The classical motion of the nuclei is determined by Newtonian physics, which amounts to solving the generalized eigenvalue problem \cite{srivastava2019physics},
\begin{equation}
    \sum_{\kappa^\prime \alpha^\prime} D_{\kappa \alpha, \kappa^\prime \alpha^\prime} (\textbf{q} ) e_{\kappa^\prime \alpha^\prime, \nu}(\textbf{q}) = \omega_{\nu, \textbf{q}}^2 e_{\kappa \alpha, \nu} (\textbf{q} ) .
\label{eigenvalue_equation}
\end{equation}
The dynamical matrix $D_{\kappa \alpha, \kappa^\prime \alpha^\prime} (\textbf{q} )$ is defined as the Fourier transform of the IFC matrix
\begin{equation}
    D_{\kappa \alpha, \kappa^\prime \alpha^\prime} (\textbf{q} ) = \frac{1}{\sqrt{m_\kappa m_{\kappa^\prime}}} \sum_{l^\prime} C_{\kappa \alpha,  \kappa^\prime \alpha^\prime}^{0, l^\prime} e^{\mathrm{i} \textbf{q} \cdot \boldsymbol{R}_l } ,
\label{IFCs}
\end{equation}
where $m_\kappa$ denotes the mass of nuclei $\kappa$, $\boldsymbol{R}_l$ is the position vector of unit cell $l$, and $\textbf{q}$ is the phonon wave vector. The solutions to Eqn. \ref{eigenvalue_equation} are the polarization vectors $e_{\kappa \alpha, \nu} (\textbf{q}) $ and frequencies $\omega_{\nu, \textbf{q}}$ corresponding to the phonon mode $\nu$ \cite{giustino2017electron, togo2015first}. 

Calculation of the phonon frequencies given by Eqn. \ref{eigenvalue_equation} on a grid of \textbf{q}-vectors allows for the determination of the phonon DOS, $\rho(\omega)$, which is defined as the number of phonon modes per unit of frequency. The total number of modes with frequencies between $\omega$ and $\omega + d\omega$ is given by $\rho(\omega) d \omega$. In a system of $N$ nuclei, there are 3$N$ modes, hence, $ \int \rho(\omega) d\omega = 3N $. After calculating the phonon modes associated with a given system it is then possible to determine the corresponding electron-phonon coupling.

\subsection{Electron-phonon matrix elements}
\label{theory_matrix_elements}

The linear electron-phonon matrix elements are computed in DFPT as \cite{giustino2007electron}
\begin{equation}
    g_{mn\nu}^{(1)} (\textbf{k}, \textbf{q}) = \sqrt{ \frac{\hbar}{2 m_0 \omega_{\nu, \textbf{q}}} } \braket{\psi_{m, \textbf{k} + \textbf{q}} | \partial_{\nu \textbf{q}} V | \psi_{n, \textbf{k}}  } ,
\end{equation}
which quantifies the coupling between the Kohn-Sham electronic states $\psi_{n, \textbf{k}}$ and $\psi_{m, \textbf{k} + \textbf{q}}$. Here, $\partial_{\nu, \textbf{q}} V$ is the change in the Kohn-Sham potential associated with a phonon of wave vector $\textbf{q}$, branch index $\nu$ with frequency $\omega_{\nu, \textbf{q}}$, $\hbar$ is the reduced Planck’s constant, and $m_0$ is the mass of the nuclei in the unit cell \cite{ponce2016epw}. 

The DFPT software we use in Sec. \ref{computational_details} has not yet implemented the ability to calculate the quadratic electron-phonon matrix elements. A widely accepted approximation \cite{plakhotnik2015electron, davies1974vibronic, maradudin1966solid} is that the quadratic elements are proportional to products of the linear elements, i.e, $g_{m n \nu \nu^\prime }^{(2)} \propto g_{m n \nu}^{(1)} \times g_{m n \nu^\prime}^{(1)}$. We assume $g_{m n \nu \nu^\prime}^{(2)} = g_{m n \nu}^{(1)} \times g_{m n \nu^\prime}^{(1)}$ and discuss in Sec. \ref{matrix_calcs} the impact that this approximation has on our results.

\section{DFT and DFPT calculations}
\label{computational_details}

\begin{figure}
	\centering
	\includegraphics[width=8.6cm]{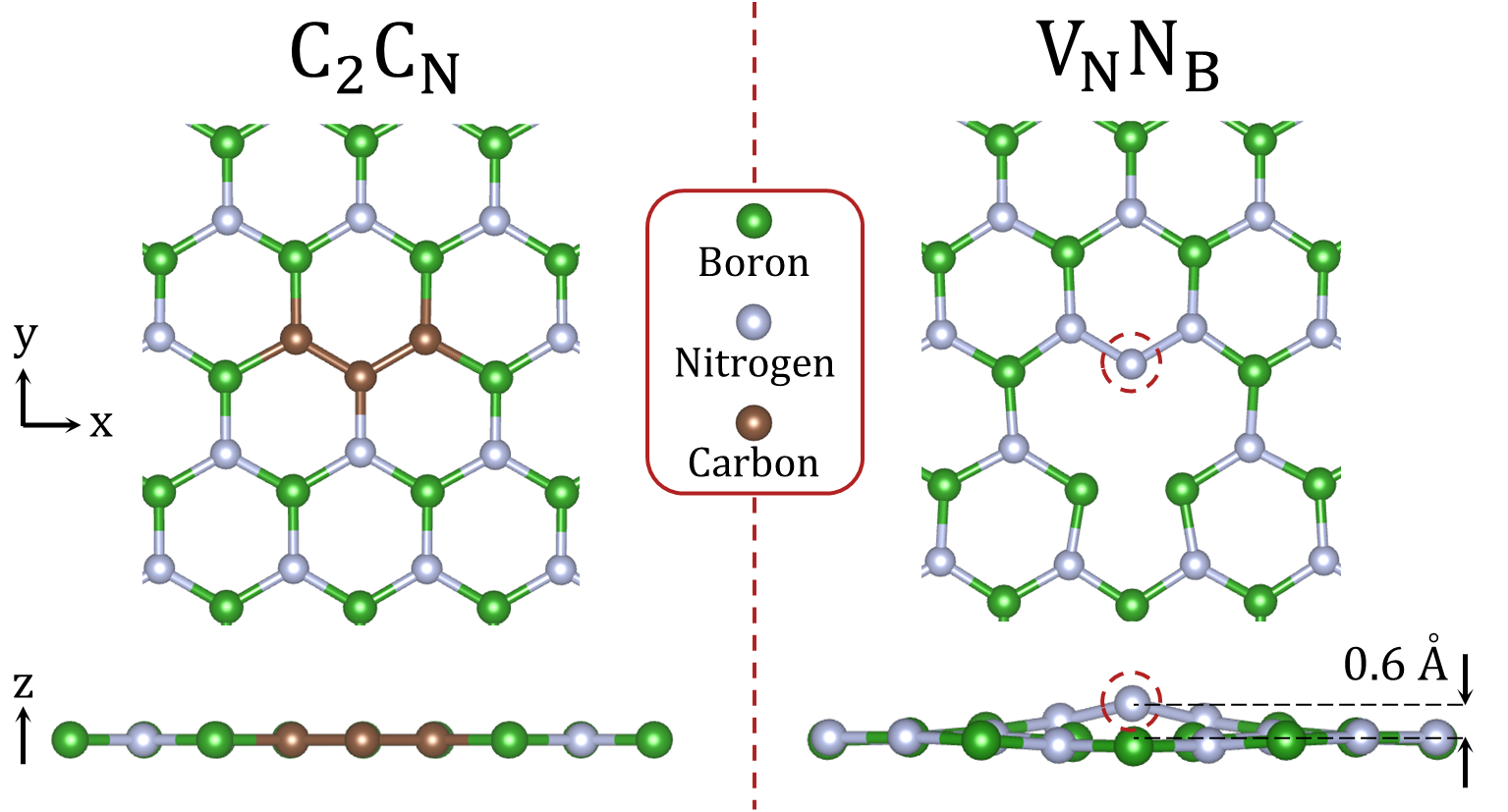}
	\caption{Relaxed atomic configurations of defects on 2D h-BN structures. The $\text{C}_2 \text{C}_\text{N}$ defect consists of adding 3 carbon atoms onto the h-BN structure as indicated on the left panel. The carbon-carbon, carbon-nitrogen, and carbon-boron bond lengths were found to be 1.41 \AA, 1.42 \AA, and 1.54 \AA, respectively. The $\text{V}_\text{N} \text{N}_\text{B}$ defect consists of making a vacancy in the h-BN structure and adding a substitutional nitrogen impurity as indicated on the right panel. The 2D h-BN layer is taken to lie on the x-y plane. Out-of-plane distortions correspond to displacements in the z-direction.}
	\label{fig:atomic_config}
\end{figure}

To shed light on the relationship between electron-phonon coupling and the transition dipole direction associated with solid-state defects in 2D materials, we investigate two defects induced on 2D h-BN structures: $\text{V}_\text{N} \text{N}_\text{B}$ \cite{abdi2018color, li2020giant} and $\text{C}_2 \text{C}_\text{N}$ \cite{golami2022b} which are depicted in Fig. \ref{fig:atomic_config}. The primary reason for selecting these defects is that our calculations reveal $\text{V}_\text{N} \text{N}_\text{B}$ and $\text{C}_2 \text{C}_\text{N}$ have out-of-plane and in-plane transition dipoles, respectively. Details of both the relaxation and the transition dipole calculations are provided in Sec. \ref{electronic_calculations}.

Additional motivation for comparing these two defects can be seen in the corresponding relaxed atomic configurations as shown in Fig. \ref{fig:atomic_config}. We find that $\text{C}_2 \text{C}_\text{N}$ lies in-plane whereas the central nitrogen in $\text{V}_\text{N} \text{N}_\text{B}$ is displaced $0.6 \text{ \AA}$ out-of-plane. The total energy of the $\text{V}_\text{N} \text{N}_\text{B}$ defect with the displaced nitrogen was found to be 37 meV lower than the case where the defect nuclei were all in-plane (see Sec. \ref{electronic_calculations} for details of the relaxation calculations). This is consistent with previous findings where defect centers belonging to the $\text{V}_\text{N} \text{X}_\text{B}$ family have an out-of-plane displacement of the $\text{X}$ impurity atom \cite{noh2018stark, li2020giant, gao2021radiative}. $\text{V}_\text{N} \text{X}_\text{B}$ defects possess a double-well potential for the out-of-plane displacement of the defect. The two wells correspond to displacements of the X impurity atom above and below the plane, and the in-plane position represents a local maximum.

The double-well potential has several characteristics that are consistent with the findings of Refs. \cite{dietrich2020solid} and \cite{hoese2020mechanical}. The first is that the strong localization of the defect orbitals around the impurity results in displacement of both the nucleus and the orbitals, possibly leading to a reduction of the coupling strength between the defect and in-plane phonon modes. The second characteristic is that the double-well potential provides an explanation as to why the emission polarization is different for resonant excitation and far off-resonance excitation. If the reflection symmetry of the defect is removed by interactions with a nearby surface or defect, then one of the wells will be lower in energy and the impurity atom will shift farther away from the plane. This can result in two sets of transitions which, in general, have different transition dipoles and polarization. It is possible that off-resonant excitations followed by nonradiative decay to one of the excited states will result in an optical transition with polarization which differs from that of the resonant transition. A defect with a double-well potential is consistent with the observations reported in Refs. \cite{jungwirth2016temperature} and \cite{dietrich2020solid}, both of which suggest the presence of a second excited state which can be excited indirectly, and has a polarization that differs from that of the resonant transition. Finally, a characteristic of a double-well potential is that various spectral properties, such as spectral diffusion, could depend on the excitation process. An excitation via a different excited state could be an explanation for different susceptibility to spectral diffusion depending on resonant or off-resonant excitation, which is consistent with experimental observations \cite{dietrich2020solid, hoese2020mechanical}.

Thus, studying the $\text{V}_\text{N} \text{N}_\text{B}$ defect allows for a simultaneous investigation of the electron-phonon coupling associated with the case of a perpendicular transition dipole moment, and the case where the central defect nucleus and the defect orbitals are displaced from the h-BN layer.

\begin{figure}
	\centering
	\includegraphics[width=8.6cm]{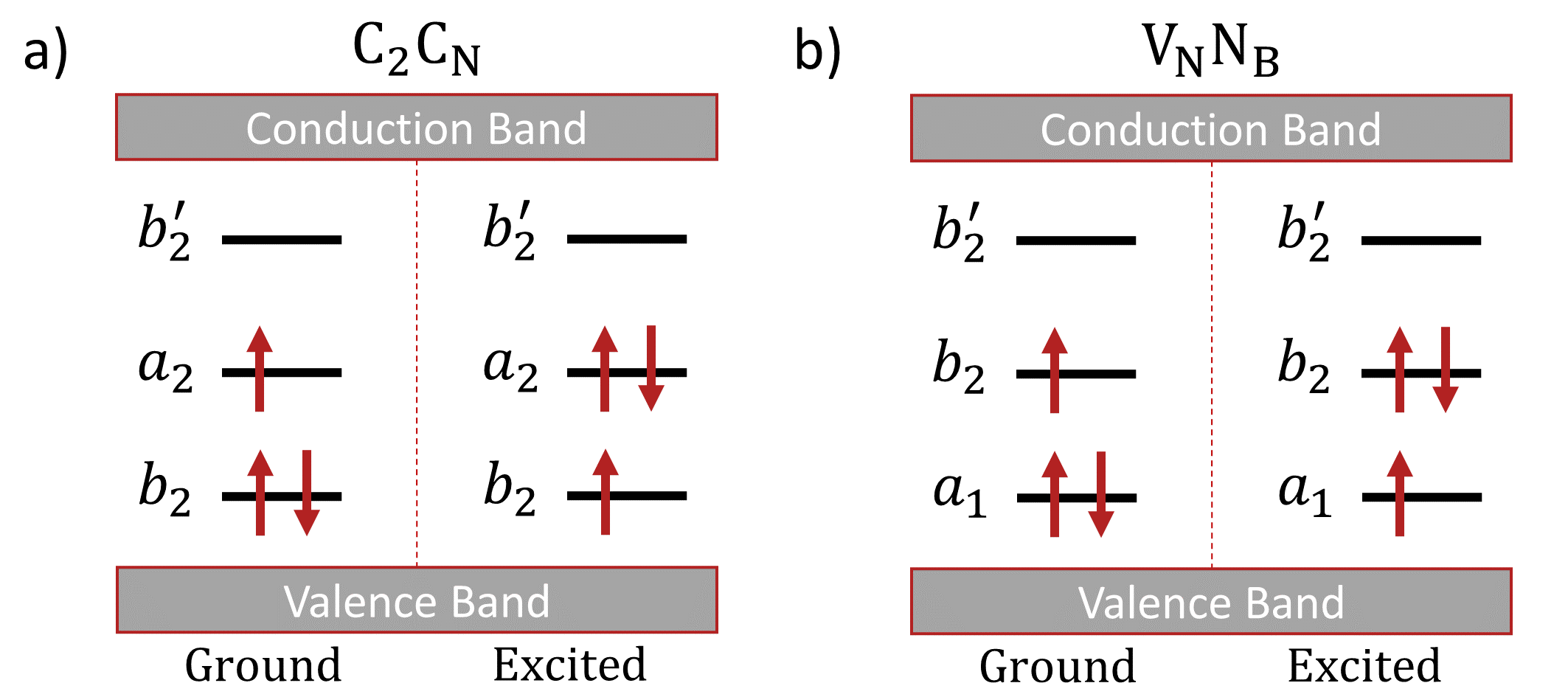}
	\caption{Single-particle defect levels and occupations corresponding to the ground and first excited electronic states of the (a) $\text{C}_2 \text{C}_\text{N}$ and (b) $\text{V}_\text{N} \text{N}_\text{B}$ defects at the $\Gamma$ point. The molecular orbitals are labeled according to their group theory symmetry-adapted orbitals \cite{abdi2018color, golami2022b}.}
	\label{fig:e_levels}
\end{figure}

The electronic energy levels of the ground and first excited states for $\text{C}_2 \text{C}_\text{N}$ and $\text{V}_\text{N} \text{N}_\text{B}$ are illustrated in Fig. \ref{fig:e_levels}. The electronic structures of these defects have some similarities. First, there are three molecular orbitals which are located within the band gap. Second, in the ground state configuration the lowest-lying level is fully occupied, the middle level is half-occupied, and the upper level is empty. Third, upon excitation, the spin-down electron in the lower level is promoted to the middle level. For both defects, we are interested in emission corresponding to the electronic transition of the spin-down electron from the middle to the lower level. Detailed investigations into the electronic structure of $\text{V}_\text{N} \text{N}_\text{B}$ and $\text{C}_2 \text{C}_\text{N}$ can be found in Refs. \cite{abdi2018color} and Ref. \cite{golami2022b}, respectively.

The \texttt{QUANTUM ESPRESSO} \cite{giannozzi2009quantum, giannozzi2017advanced, giannozzi2020quantum} software package is a well-established computational toolkit for predicting electronic and phonon properties of defects in solid-state systems. The following DFT and DFPT \cite{giustino2007electron, ponce2016epw, noffsinger2010epw} calculations and post-processing were performed using \texttt{QUANTUM ESPRESSO}. Various assumptions and simplifications made in our analysis are specific to the current capabilities offered by this package.

\subsection{Electronic structure calculations}
\label{electronic_calculations}

Our DFT calculations in \texttt{QUANTUM ESPRESSO} were performed using the generalized gradient approximation (GGA), with the Perdew-Burke-Ernzerhof (PBE) functional \cite{perdew1996generalized} for the electron exchange and correlation potentials. We used a plane-wave basis set with a wave function kinetic energy cutoff of 60 Ry and projector augmented wave (PAW) pseudopotentials \cite{blochl1994projector}. The total energy was found to converge at approximately 40 Ry, however, the larger cutoff was selected to ensure the accurate calculation of the phonon modes.

Our super-cells consisted of 50 atoms corresponding to $5 \times 5$ unit cells of mono-layer h-BN. The structures were relaxed until the force on each atom was less than $1 \times 10^{-6}$ Ry/au, and we used a convergence threshold for the self consistency calculations of $1\times 10^{-12}$ Ry. Relaxation was performed by allowing the in-plane lattice vectors to relax, and the out-of-plane lattice vector was held fixed to impose a vacuum spacing of 15 \AA. The electronic and phonon calculations were run within the 2D framework offered by \texttt{QUANTUM ESPRESSO}, where the Coulomb interaction is truncated in the z-direction \cite{sohier2017density}. Other works that also inspired the settings of our DFT calculations and serve as important references for the calibration of the pristine h-BN calculations and the defect study cases are Refs. \cite{leon2019interface, satawara2021structural, miro2014atlas, angizi2020two, huang2022carbon, zhang2020point, kim2017geometric, liu2003structural}.

The total energy converged with respect to the Brillouin zone sampling density, and the lower-limit necessary to achieve convergence, a $2 \times 2 \times 1$ \textbf{k}-grid, was used. Both the super-cell size and the sampling density are important factors which dictate the run-time associated with the DFPT calculations. In particular, the matrix element calculations are computationally expensive in terms of run-time. The use of larger super-cells and/or more dense $ \textbf{k} $-grids resulted in calculations that were infeasible. Thus, the first step was to verify that our 50-atom super-cell calculations managed to capture several key properties of the defect that are of interest in our study.

\begin{figure}
	\centering
	\includegraphics[width=8.6cm]{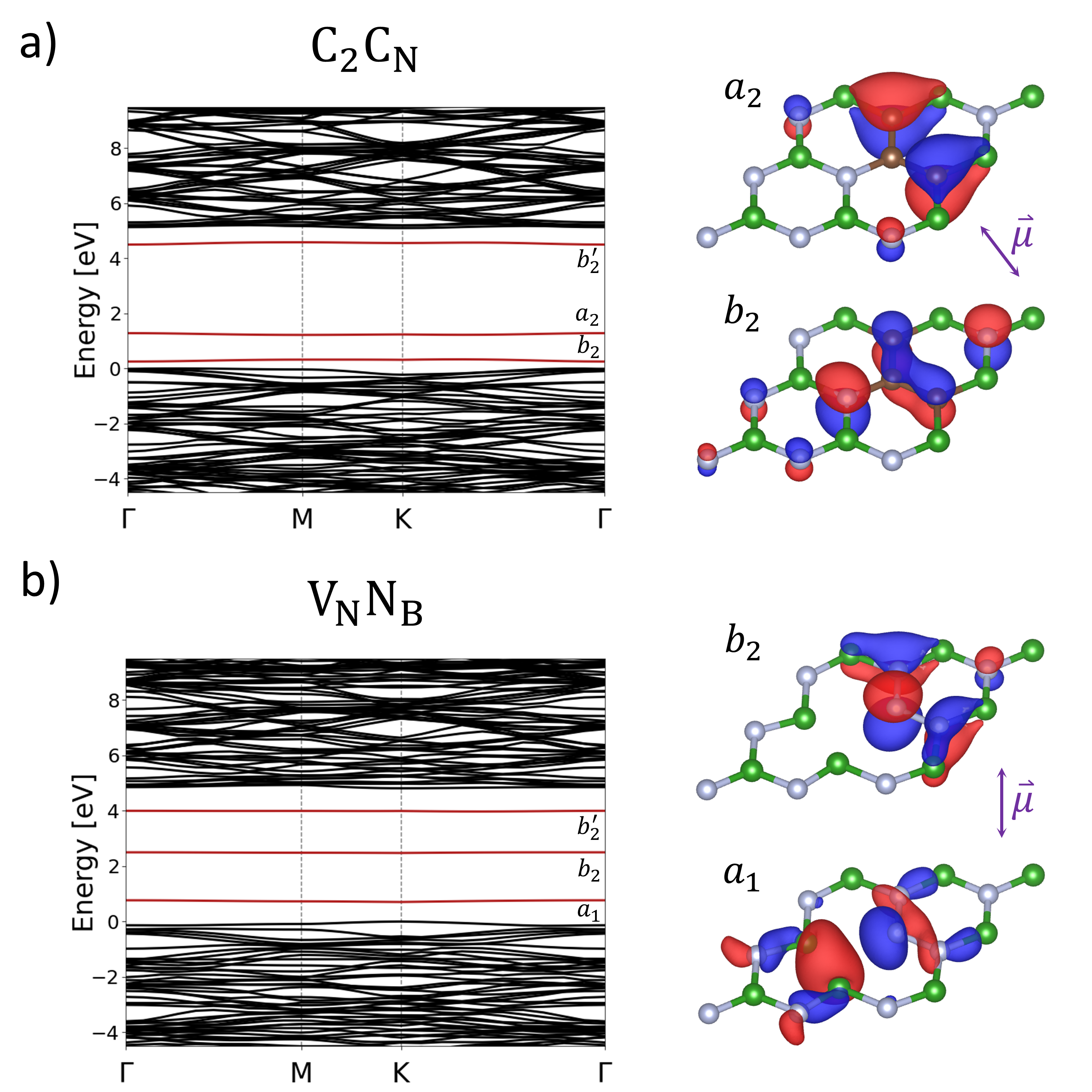}
	\caption{Electronic band structure and molecular orbitals for the (a) $\text{C}_2 \text{C}_\text{N}$ and (b) $\text{V}_\text{N} \text{N}_\text{B}$ defect structures. Relatively low dispersion of the defect states was obtained for a 50 atom super-cell, corresponding to $5 \times 5$ unit cells of h-BN.}
	\label{fig:e_bands}
\end{figure}

In Fig. \ref{fig:e_bands}, we plot the electronic band structure and wave functions obtained for both defects. Our calculations yield three defect states within the energy gap as expected, and inspection of the wave functions reveal that the relative ordering of the states is the same as obtained in Refs. \cite{abdi2018color} and \cite{golami2022b}. Additionally, the defect states are relatively flat, that is, the energy does not vary significantly along the Brillouin zone path. We found that smaller super-cell models resulted in dispersive bands across the Brillouin zone, which is due to defect-defect interactions resulting from the periodic replicas of the super-cell.

Given that the main features of the electronic band structure and defect wave functions associated with the 50-atom cell are consistent with the findings of Refs. \cite{abdi2018color} and \cite{golami2022b}, we conclude that the strong localization of the defect wave functions around the defect nuclei makes it possible to provide a reasonable description of the electronic structure using a medium size super-cell.

The two defects were initially of interest to us because inspection of the molecular orbitals, following the method presented in Ref. \cite{walla2014modern}, indicated that the transition dipole moment associated with the first excited state to ground state transition of $\text{C}_2 \text{C}_\text{N}$ would lie in the h-BN plane (taken to be the x-y plane), and that of $\text{V}_\text{N} \text{N}_\text{B}$ would have transition dipole perpendicular to the plane. Inspection of Fig. \ref{fig:e_bands} reveals this to be the case since the molecular orbitals of $\text{C}_2 \text{C}_\text{N}$ are both antisymmetric with respect to in-plane mirror symmetry, resulting in a transition dipole with a z-component equal to zero. On the other hand, the $b_2$ and $a_1$ orbitals of $\text{V}_\text{N} \text{N}_\text{B}$ are antisymmetric and symmetric, respectively, which indicates that the z-component of the transition dipole can be nonzero.

To quantify the direction of the transition dipoles, the \texttt{WFCK2R.X} module in \texttt{QUANTUM ESPRESSO \cite{giannozzi2009quantum, giannozzi2017advanced, giannozzi2020quantum}} was used to produce the real-space wave functions, from which the transition dipole moments were directly computed according to
\begin{equation}
    \boldsymbol{\mu} = \braket{\psi_f | e \boldsymbol{r} | \psi_i},
\end{equation} where $\psi_i$ and $\psi_f$ are the initial and final wave functions, $e$ is the elementary charge, and $\boldsymbol{r}$ is the real-space unit vector. The results of the calculations are presented in Table \ref{transition_dipole_table}.

\begin{table}[h]
\centering
\caption{Calculated transition dipole moment between the ground and first excited states of the $\text{C}_2 \text{C}_\text{N}$ and $\text{V}_\text{N} \text{N}_\text{B}$ defects.}
\begin{tabular}{>{\centering\arraybackslash}p{1.0cm} >{\centering\arraybackslash}p{1.6cm} >{\centering\arraybackslash}p{1.8cm} >{\centering\arraybackslash}p{1.8cm} >{\centering\arraybackslash}p{1.8cm}} 
& & & & \\[-1ex]
\hline \hline
& & & & \\[-1.5ex]
Defect & $\boldsymbol{\mu}$ & $\mu$ [e\AA] &  $\sqrt{\mu_x^2 + \mu_y^2} / \mu$ &  $\mu_z / \mu$ \\
& & & & \\[-1.5ex]
\hline
& & & & \\[-1.5ex]
$\text{C}_2 \text{C}_\text{N}$ & $\braket{b_2|e \boldsymbol{r}|a_2}$ &  $0.93$ & $1.0$ &  $1.3 \times 10^{-8}$ \\
$\text{V}_\text{N} \text{N}_\text{B}$ & $\braket{a_1|e \boldsymbol{r}|b_2}$ & $2.5 \times 10^{-3}$ & $ 2.4 \times 10^{-6} $ & $1.0$ \\ 
& & & & \\[-1.5ex]
\hline \hline
\end{tabular}
\label{transition_dipole_table}
\end{table}

The transition dipole associated with $\text{V}_\text{N} \text{N}_\text{B}$ is significantly smaller than that of $\text{C}_2 \text{C}_\text{N}$. The reason for this is that the $\text{C}_2 \text{C}_\text{N}$ charge density in both the ground and excited states is much more localized to the carbon atoms, as compared to the ground state charge density of $\text{V}_\text{N} \text{N}_\text{B}$ which is spread around the vacancy. 

The last column in Table \ref{transition_dipole_table} presents the ratio of the magnitude of the out-of-plane component of the transition dipole to the total magnitude. Consistent with the graphical estimation, the transition dipole associated with the ZPL transition of $\text{C}_2 \text{C}_\text{N}$ is found to lie in-plane, and that of $\text{V}_\text{N} \text{N}_\text{B}$ is perpendicular to the plane.

\begin{figure*}
	\centering
	\includegraphics[width=\textwidth]{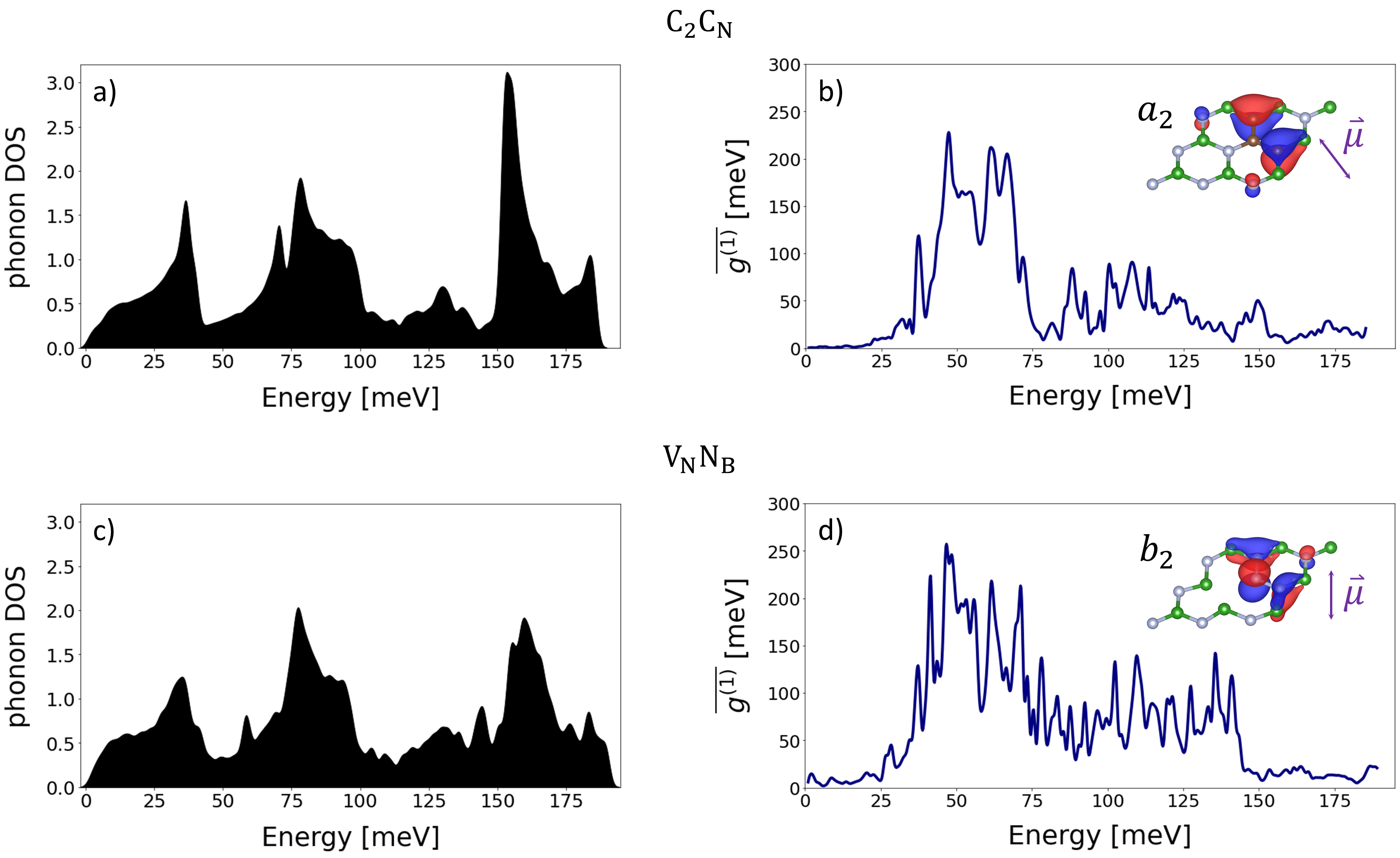}
	\caption{Phonon DOS of the (a) $\text{C}_2 \text{C}_\text{N}$ and (c) $\text{V}_\text{N} \text{N}_\text{B}$ defects in mono-layer h-BN. Average electron-phonon coupling strengths as a function of energy associated with intraband scattering processes which occur in the excited electronic state of the (b) $\text{C}_2 \text{C}_\text{N}$ and (d) $\text{V}_\text{N} \text{N}_\text{B}$ defect structures.}
	\label{fig:spectral_density}
\end{figure*}

\subsection{Phonon DOS}
\label{phnon_dos}

The model which successfully describes the ZPL broadening of the NV$^-$ center \cite{plakhotnik2015electron} considers electron-phonon interaction rates occurring in the excited electronic state. When the electron associated with a transition is promoted to a higher-energy orbital, the nuclear configuration shifts in response to the change in electron density. While it is possible to perform DFT calculations in the excited-state nuclear configuration using the $\Delta$SCF method \cite{gali2009theory}, it is not yet possible using \texttt{QUANTUM ESPRESSO} to perform the DFPT calculations outside of the ground-state nuclear configuration. As such, we assume that the calculations performed in the ground-state nuclear configuration of both the phonon DOS and the electron-phonon couplings in the excited electronic state, provide an adequate description of the corresponding quantities in the excited nuclear configuration. This assumption represents a notable limitation in our analysis, and further discussion is provided in Sec. \ref{discussion}. 

The phonon DOS was computed following several steps. First, the \texttt{PH.X} module in \texttt{QUANTUM ESPRESSO} \cite{giannozzi2009quantum, giannozzi2017advanced, giannozzi2020quantum} was used to perform the DFPT calculations and compute the dynamical matrices using a $2\times 2 \times 1$ $\textbf{q}$-grid, which is commensurate with the electronic grid as required by the matrix element calculations. The convergence threshold was set to $1\times 10^{-16}$ Ry. 

Calculation of the dynamical matrices was split-up into four separate calculations using the same general methodology as outlined in Ref. \cite{di2009calculation}. While the 50-atom super-cell is sufficiently small to run the DFPT calculation as a single job over several days,  depending on the computational resources available, it may be useful to manage this calculation in parts, i.e., by dividing the calculation into many sub-tasks to efficiently schedule and process the results.

Next, the real-space IFCs were calculated from the dynamical matrices using the \texttt{QUANTUM ESPRESSO} \texttt{Q2R.X} module \cite{giannozzi2009quantum, giannozzi2017advanced, giannozzi2020quantum} with the simple acoustic-sum-rule imposed. Finally, the \texttt{QUANTUM ESPRESSO} \texttt{MATDYN.X} module \cite{giannozzi2009quantum, giannozzi2017advanced, giannozzi2020quantum} was used to calculate phonon frequencies at generic 
$\textbf{q}$-points, via Fourier transform of the IFCs as described by Eqn. \ref{IFCs}. The phonon DOS was calculated on a dense $50\times 50 \times 1$ grid of $\textbf{q}$-points. Gaussian smoothing was then applied using a Full Width at Half Maximum of 3 meV (0.725 THz). An additional observation made in Ref. \cite{hoese2020mechanical} which we have not yet discussed is that all emitters which produce narrow lines have one-phonon bands with a gap of approximately 2 THz in the low-energy regime, indicating a reduction of either the phonon density or the electron-phonon coupling associated with low-energy acoustic phonons. The Gaussian smoothing was chosen such that the resolution is high enough to enable a sampling rate capable of resolving low-energy gaps of approximately 2 THz. The resulting phonon DOS for both defects are shown in Fig. \ref{fig:spectral_density}.

Comparing the DOS of both defects, there are no features which immediately suggest that the phonon densities are suppressed in $\text{V}_\text{N} \text{N}_\text{B}$. According to the theory of Ref. \cite{plakhotnik2015electron}, contributions from low-energy acoustic phonon modes have the dominant impact on ZPL broadening. The peak at approximately 40 meV is likely to be the acoustic mode cutoff energy (the highest possible energy of the acoustic modes). Both defects have non-negligible contributions to the DOS from phonons with energies below the cutoff, indicating that if the direction of the transition dipole does have an significant effect on the electron-phonon interaction rates, then such a mechanism must manifest itself in the matrix elements.

\subsection{Matrix elements}
\label{matrix_calcs}

Calculation of the electron-phonon matrix elements from first-principles is challenging because of the necessity of sampling the Brillouin zone using a very dense grid of \textbf{q}-points. The \texttt{EPW} software package \cite{giustino2007electron, ponce2016epw, noffsinger2010epw}, which is fully integrated in and distributed with the \texttt{QUANTUM ESPRESSO} package, provides the capability to calculate the linear matrix element coupling terms in Eqn. \ref{Hamiltonian}. The recent release of \texttt{EPW v5.4} (7/2021) has implemented the 2D framework \cite{sohier2017density} used in the electronic and phonon calculations, making it possible to calculate the coupling in 2D materials.

\texttt{EPW} exploits the real-space localization of Wannier functions to generate large numbers of electron–phonon matrix elements through a generalized Fourier interpolation. The matrix elements are computed on a coarse grid of electron and phonon wave vectors in the Bloch representation. They are then transformed from the Bloch representation to the maximally localized Wannier functions representation. Finally, they are Fourier transformed back to the Bloch representation on arbitrarily dense grids of electron and phonon wave vectors.

After performing the electronic and DFPT calculations described in Sec. \ref{electronic_calculations} and \ref{phnon_dos}, the matrix elements were computed on a dense grid of $100\times 100 \times 1$ $\textbf{q}$-points at the $\Gamma$ $\textbf{k}$-point. The resulting matrix elements were ordered according to phonon energy, partitioned into 1 meV bins, and the average within each bin was computed.

As previously discussed, our focus is on the coupling associated with intraband electronic transitions in the excited electronic state. For $\text{C}_2 \text{C}_\text{N}$ the excited electronic band is $a_2$, and for $\text{V}_\text{N} \text{N}_\text{B}$ it is the $b_2$ band (see Fig. \ref{fig:e_bands}). The calculated average electron-phonon coupling strengths for the intraband processes were plotted as a function of energy and the results of both defect structures are shown in Fig. \ref{fig:spectral_density}. While the low-energy regime is of primary interest to our analysis, our calculations are not able to resolve the fine features associated with the acoustic phonons. We performed the same calculations for smaller super-cells and there does not appear to be convergence of the features in the low-energy regime. As such, we have plotted the entire range of phonon energies rather than a magnified view of the acoustic regime.

The first-order elastic scattering processes which involve the absorption of one phonon and the emission of another phonon with the same energy occur at rates that depend on the quadratic matrix elements. As mentioned in Sec. \ref{theory_matrix_elements}, we approximate the quadratic elements as products of the linear elements. For intraband mechanisms, the quantity of interest is then the square of the linear elements, meaning it is sufficient to compare the magnitudes of the linear elements of the two defects.

As with the case of the phonon DOS, there are obvious differences between the results of $\text{C}_2 \text{C}_\text{N}$ and $\text{V}_\text{N} \text{N}_\text{B}$, however, there is no clear reduction in the strength of the electron-phonon coupling in the low energy acoustic mode regime. In fact, the matrix elements associated with the perpendicular transition dipole are slightly larger than the case of the parallel dipole in the acoustic regime. The matrix elements were calculated at several $\textbf{k}$-points in addition to the $\Gamma$ point and the corresponding results show negligible differences when compared to Fig. \ref{fig:spectral_density}.

In order to exclude the theoretical possibility that the coupling might be low for both defects, we compare the matrix elements to those associated with the NV$^-$ center in diamond (K. Sharman \textit{et al.}, in preparation). For phonon energies less than 20 meV, the matrix elements associated with the $\text{C}_2 \text{C}_\text{N}$ and $\text{V}_\text{N} \text{N}_\text{B}$ defects are approximately an order of magnitude greater than those associated with the NV$^-$ center, and up to two orders of magnitude greater at energies above 20 meV. Further, our DFT calculations indicate that the DOS associated with the NV$^-$ center is less concentrated at lower energies as compared to the two defects studied here.  Since the NV$^-$ center is known to have ZPL widths on the order of $10^3$ GHz at room temperature \cite{plakhotnik2015electron}, the fact that the defects studied here have larger matrix elements and DOS at low phonon energies as compared to the NV$^-$ center suggests that it would be unreasonable to expect that either of these two defects would exhibit FT-limited linewidths at room temperature.

The lack of a clear decoupling in either the DOS or the matrix elements leads to the conclusion that the direction of the transition dipole alone is not sufficient to result in an appreciable lowering of the electron-phonon coupling associated with defects in 2D materials. Similarly, the out-of-plane displacement of the central nitrogen atom in $\text{V}_\text{N} \text{N}_\text{B}$ is not expected to be, by itself, the mechanism responsible for the FT-limited lines reported in Refs. \cite{dietrich2020solid} and \cite{hoese2020mechanical}. Finally, we see no indication that the orientation of the transition dipole can explain the low-energy gap in the one-phonon bands presented in Ref. \cite{hoese2020mechanical}.

It is worth mentioning that our analysis of the electron-phonon coupling is not directly related to the phonon sideband (PSB). The calculated couplings presented here correspond to intraband electronic transitions. The PSB on the other hand, involves interband transitions, namely, transitions between the excited and ground states. An interesting future project would be to use related DFT calculations to calculate the PSB of a given defect. We expect that for defects in h-BN, we should see relatively large couplings between the excited and ground states at approximately 160 meV since this feature appears in many of the experimental observations \cite{jara2021first}.


\section{Discussion}
\label{discussion}

The results presented in the preceding section are subject to a number of assumptions and limitations. We suspect that the 50-atom super-cells used to model the defects represent that the largest shortcoming associated with our analysis. The electronic calculations revealed that periodic replicas of the defect are sufficiently far from one another, however, we expect that the phonon modes would be somewhat different if the size of the super-cell was increased. In any case, our calculations serve as a stepping stone to enable finite-size supercell corrections to be implemented that can further improve these findings quantitatively. Examples of such finite-size correction methods that can be tested in electronic structure calculations have been vastly published, e.g., \cite{rocha2018finite, castleton2004finite, castleton2006managing, freysoldt2009fully, komsa2012finite, kwee2008finite, yoo2021finite}. Similar methods  based on extrapolation schemes could be used to analyze finite-size effects in the phonon modes.

The low sampling density of both the $\textbf{k}$- and $\textbf{q}$-points is another significant limitation in our analysis. Fourier interpolation allows for calculation of both the phonon DOS and the matrix elements on fine grids, however, it is possible that subtle details are not adequately captured with the coarse grids. Along the same lines, our calculations are not able to resolve the mode-specific contributions to either the matrix elements or the phonon DOS. It is possible that the in-plane acoustic mode is indeed decoupled from the electronic state, but this information is hidden in our analysis which considers the coupling of all modes.

The electronic calculations did not utilize the Heyd-Scuseria-Ernzerhof (HSE) hybrid functional \cite{heyd2003hybrid}, which has been shown to provide an accurate estimate of the band gap of h-BN, because the DFPT calculations implemented in \texttt{QUANTUM ESPRESSO} do not support the use of hybrid functionals. The impact of not using the HSE functional can be seen in the underestimation of the energy spacing between defect states, and by extension, an underestimation of the ZPL energy. The matrix element calculations do not consider interband transitions, as discussed in Sec. \ref{theory_electron_phonon}, and therefore, underestimating the energy spacing between electronic states is expected to have negligible impact on our results.

Additionally, the electronic and phonon calculations were performed in the ground-state nuclear configuration. A more accurate description of the excited electronic state and the corresponding phonon modes could be provided by using spin-polarized, constrained occupancy calculations with the HSE functional. We suggest that future software developments allow such calculations to be run when determining both the phonon DOS and the electron-phonon coupling. When available, the electron-phonon coupling calculated in the excited state framework can be compared to that of the ground state framework presented here.

An extension of our analysis would be to calculate the ZPL width as a function of temperature. This is an interesting route, not only to identify the defects with narrow lines at room temperature, but also to identify solid-state defects in general which have unknown atomic structures. As demonstrated by the NV$^-$ center and the silicon-vacancy center in diamond, defects in the same host material can have very different electron-phonon interactions, and the ZPL widths have different temperature dependencies. Thus, the ability to calculate the ZPL broadening would allow researchers to make additional theoretical predictions which could then be compared to experimental observations.

To calculate the ZPL width as a function of temperature, one would need the ability to directly calculate the quadratic electron-phonon matrix elements, determine the acoustic mode cutoff energies, as well as the symmetry-specific mode contributions to the phonon DOS \cite{plakhotnik2015electron}. Improvements made in these areas would represent a step closer towards a complete description of the line broadening mechanisms from first-principles. 

\section{Conclusion}
\label{conclusion}

The discovery of quantum emitters in layered h-BN structures that are decoupled from in-plane phonons reveals that defects in solid-state systems have the potential to serve as indistinguishable SPSs which can operate at room temperature. The underlying mechanism responsible for the decoupling remains a mystery, however, the leading hypothesis is that an out-of-plane distortion of the emitter's orbitals leads to a reduction of the electron-phonon coupling. Here, we have tested the theory that the direction of the transition dipole moment will directly impact the rate at which electron-phonon interactions occur. Our DFT calculations show that the transition dipole of $\text{C}_2 \text{C}_\text{N}$ is in-plane, and that of $\text{V}_\text{N} \text{N}_\text{B}$ is out-of-plane. DFPT calculations were used to investigate and compare the phonon DOS and the electron-phonon coupling associated with these two defects. 

Our results show no clear indication that the electron-phonon interactions are drastically different for the two defects, suggesting that an out-of-plane transition dipole moment in 2D h-BN is not sufficient to suppress the mechanisms responsible for temperature-dependent ZPL broadening. Our findings also indicate that the out-of-plane displacement of the nitrogen impurity atom in $\text{V}_\text{N} \text{N}_\text{B}$ does not lead to an obvious reduction in the electron-phonon scattering processes.

We have discussed the various limitations associated with our calculations, in hopes that future software developments can address these issues. Our detailed description of calculating the electron-phonon coupling will be of interest in the field of solid-state quantum information processing, and adds to the ever-growing toolbox of DFT methods available to researchers.

Since all mechanically decoupled emitters were found in bulk-like h-BN, the multi-layer nature of the host material could be an essential ingredient for the decoupling mechanism \cite{kubanek2022review}. The many layers in bulk-like h-BN could provide a trap, for example, for molecules localized between the layers. In such a scenario, the acoustic coupling to in-plane phonon modes could be suppressed while optical coupling would still be present in agreement with experimental indications. Investigation of this hypothesis represents an interesting future research direction.

\section*{Acknowledgments}
The authors thank Jordan Smith for useful discussions. This work was supported by the Natural Sciences and Engineering Research Council of Canada (NSERC) through its Discovery Grant, Canadian Graduate Scholarships, CREATE, and Strategic Project Grant programs; by Alberta Innovates through its Graduate Student Scholarship program, and by the Alberta government through its Graduate Excellence Scholarship (AGES) program. We also acknowledge the Advanced Research Computing (ARC) IT team of the University of Calgary, and the Digital Research Alliance of Canada (\href{https://alliancecan.ca/en}{alliancecan.ca/en}) for computational resources.

\bibliographystyle{apsrev4-1}
\bibliography{ref}

\begin{thebibliography}{87}%
\makeatletter
\providecommand \@ifxundefined [1]{%
 \@ifx{#1\undefined}
}%
\providecommand \@ifnum [1]{%
 \ifnum #1\expandafter \@firstoftwo
 \else \expandafter \@secondoftwo
 \fi
}%
\providecommand \@ifx [1]{%
 \ifx #1\expandafter \@firstoftwo
 \else \expandafter \@secondoftwo
 \fi
}%
\providecommand \natexlab [1]{#1}%
\providecommand \enquote  [1]{``#1''}%
\providecommand \bibnamefont  [1]{#1}%
\providecommand \bibfnamefont [1]{#1}%
\providecommand \citenamefont [1]{#1}%
\providecommand \href@noop [0]{\@secondoftwo}%
\providecommand \href [0]{\begingroup \@sanitize@url \@href}%
\providecommand \@href[1]{\@@startlink{#1}\@@href}%
\providecommand \@@href[1]{\endgroup#1\@@endlink}%
\providecommand \@sanitize@url [0]{\catcode `\\12\catcode `\$12\catcode
  `\&12\catcode `\#12\catcode `\^12\catcode `\_12\catcode `\%12\relax}%
\providecommand \@@startlink[1]{}%
\providecommand \@@endlink[0]{}%
\providecommand \url  [0]{\begingroup\@sanitize@url \@url }%
\providecommand \@url [1]{\endgroup\@href {#1}{\urlprefix }}%
\providecommand \urlprefix  [0]{URL }%
\providecommand \Eprint [0]{\href }%
\providecommand \doibase [0]{http://dx.doi.org/}%
\providecommand \selectlanguage [0]{\@gobble}%
\providecommand \bibinfo  [0]{\@secondoftwo}%
\providecommand \bibfield  [0]{\@secondoftwo}%
\providecommand \translation [1]{[#1]}%
\providecommand \BibitemOpen [0]{}%
\providecommand \bibitemStop [0]{}%
\providecommand \bibitemNoStop [0]{.\EOS\space}%
\providecommand \EOS [0]{\spacefactor3000\relax}%
\providecommand \BibitemShut  [1]{\csname bibitem#1\endcsname}%
\let\auto@bib@innerbib\@empty
\bibitem [{\citenamefont {Knill}\ \emph {et~al.}(2001)\citenamefont {Knill},
  \citenamefont {Laflamme},\ and\ \citenamefont {Milburn}}]{knill2001scheme}%
  \BibitemOpen
  \bibfield  {author} {\bibinfo {author} {\bibfnamefont {E.}~\bibnamefont
  {Knill}}, \bibinfo {author} {\bibfnamefont {R.}~\bibnamefont {Laflamme}}, \
  and\ \bibinfo {author} {\bibfnamefont {G.~J.}\ \bibnamefont {Milburn}},\
  }\href {https://www.nature.com/articles/35051009} {\bibfield  {journal}
  {\bibinfo  {journal} {Nature}\ }\textbf {\bibinfo {volume} {409}},\ \bibinfo
  {pages} {46} (\bibinfo {year} {2001})}\BibitemShut {NoStop}%
\bibitem [{\citenamefont {Kok}\ \emph {et~al.}(2007)\citenamefont {Kok},
  \citenamefont {Munro}, \citenamefont {Nemoto}, \citenamefont {Ralph},
  \citenamefont {Dowling},\ and\ \citenamefont {Milburn}}]{kok2007linear}%
  \BibitemOpen
  \bibfield  {author} {\bibinfo {author} {\bibfnamefont {P.}~\bibnamefont
  {Kok}}, \bibinfo {author} {\bibfnamefont {W.~J.}\ \bibnamefont {Munro}},
  \bibinfo {author} {\bibfnamefont {K.}~\bibnamefont {Nemoto}}, \bibinfo
  {author} {\bibfnamefont {T.~C.}\ \bibnamefont {Ralph}}, \bibinfo {author}
  {\bibfnamefont {J.~P.}\ \bibnamefont {Dowling}}, \ and\ \bibinfo {author}
  {\bibfnamefont {G.~J.}\ \bibnamefont {Milburn}},\ }\href
  {https://journals.aps.org/rmp/abstract/10.1103/RevModPhys.79.135} {\bibfield
  {journal} {\bibinfo  {journal} {Reviews of Modern Physics}\ }\textbf
  {\bibinfo {volume} {79}},\ \bibinfo {pages} {135} (\bibinfo {year}
  {2007})}\BibitemShut {NoStop}%
\bibitem [{\citenamefont {Jennewein}\ \emph {et~al.}(2000)\citenamefont
  {Jennewein}, \citenamefont {Achleitner}, \citenamefont {Weihs}, \citenamefont
  {Weinfurter},\ and\ \citenamefont {Zeilinger}}]{jennewein2000fast}%
  \BibitemOpen
  \bibfield  {author} {\bibinfo {author} {\bibfnamefont {T.}~\bibnamefont
  {Jennewein}}, \bibinfo {author} {\bibfnamefont {U.}~\bibnamefont
  {Achleitner}}, \bibinfo {author} {\bibfnamefont {G.}~\bibnamefont {Weihs}},
  \bibinfo {author} {\bibfnamefont {H.}~\bibnamefont {Weinfurter}}, \ and\
  \bibinfo {author} {\bibfnamefont {A.}~\bibnamefont {Zeilinger}},\ }\href
  {https://aip.scitation.org/doi/abs/10.1063/1.1150518} {\bibfield  {journal}
  {\bibinfo  {journal} {Review of Scientific Instruments}\ }\textbf {\bibinfo
  {volume} {71}},\ \bibinfo {pages} {1675} (\bibinfo {year}
  {2000})}\BibitemShut {NoStop}%
\bibitem [{\citenamefont {Hoese}\ \emph {et~al.}(2022)\citenamefont {Hoese},
  \citenamefont {Koch}, \citenamefont {Breuning}, \citenamefont {Lettner},
  \citenamefont {Fehler},\ and\ \citenamefont {Kubanek}}]{hoese2022single}%
  \BibitemOpen
  \bibfield  {author} {\bibinfo {author} {\bibfnamefont {M.}~\bibnamefont
  {Hoese}}, \bibinfo {author} {\bibfnamefont {M.~K.}\ \bibnamefont {Koch}},
  \bibinfo {author} {\bibfnamefont {F.}~\bibnamefont {Breuning}}, \bibinfo
  {author} {\bibfnamefont {N.}~\bibnamefont {Lettner}}, \bibinfo {author}
  {\bibfnamefont {K.~G.}\ \bibnamefont {Fehler}}, \ and\ \bibinfo {author}
  {\bibfnamefont {A.}~\bibnamefont {Kubanek}},\ }\href
  {https://aip.scitation.org/doi/full/10.1063/5.0074946} {\bibfield  {journal}
  {\bibinfo  {journal} {Applied Physics Letters}\ }\textbf {\bibinfo {volume}
  {120}},\ \bibinfo {pages} {044001} (\bibinfo {year} {2022})}\BibitemShut
  {NoStop}%
\bibitem [{\citenamefont {Gisin}\ and\ \citenamefont
  {Thew}(2007)}]{gisin2007quantum}%
  \BibitemOpen
  \bibfield  {author} {\bibinfo {author} {\bibfnamefont {N.}~\bibnamefont
  {Gisin}}\ and\ \bibinfo {author} {\bibfnamefont {R.}~\bibnamefont {Thew}},\
  }\href {https://www.nature.com/articles/nphoton.2007.22} {\bibfield
  {journal} {\bibinfo  {journal} {Nature Photonics}\ }\textbf {\bibinfo
  {volume} {1}},\ \bibinfo {pages} {165} (\bibinfo {year} {2007})}\BibitemShut
  {NoStop}%
\bibitem [{\citenamefont {Sangouard}\ \emph {et~al.}(2011)\citenamefont
  {Sangouard}, \citenamefont {Simon}, \citenamefont {De~Riedmatten},\ and\
  \citenamefont {Gisin}}]{sangouard2011quantum}%
  \BibitemOpen
  \bibfield  {author} {\bibinfo {author} {\bibfnamefont {N.}~\bibnamefont
  {Sangouard}}, \bibinfo {author} {\bibfnamefont {C.}~\bibnamefont {Simon}},
  \bibinfo {author} {\bibfnamefont {H.}~\bibnamefont {De~Riedmatten}}, \ and\
  \bibinfo {author} {\bibfnamefont {N.}~\bibnamefont {Gisin}},\ }\href
  {https://journals.aps.org/rmp/abstract/10.1103/RevModPhys.83.33} {\bibfield
  {journal} {\bibinfo  {journal} {Reviews of Modern Physics}\ }\textbf
  {\bibinfo {volume} {83}},\ \bibinfo {pages} {33} (\bibinfo {year}
  {2011})}\BibitemShut {NoStop}%
\bibitem [{\citenamefont {Wu}\ \emph {et~al.}(2020)\citenamefont {Wu},
  \citenamefont {Liu},\ and\ \citenamefont {Simon}}]{wu2020near}%
  \BibitemOpen
  \bibfield  {author} {\bibinfo {author} {\bibfnamefont {Y.}~\bibnamefont
  {Wu}}, \bibinfo {author} {\bibfnamefont {J.}~\bibnamefont {Liu}}, \ and\
  \bibinfo {author} {\bibfnamefont {C.}~\bibnamefont {Simon}},\ }\href
  {https://journals.aps.org/pra/abstract/10.1103/PhysRevA.101.042301}
  {\bibfield  {journal} {\bibinfo  {journal} {Physical Review A}\ }\textbf
  {\bibinfo {volume} {101}},\ \bibinfo {pages} {042301} (\bibinfo {year}
  {2020})}\BibitemShut {NoStop}%
\bibitem [{\citenamefont {Somaschi}\ \emph {et~al.}(2016)\citenamefont
  {Somaschi}, \citenamefont {Giesz}, \citenamefont {De~Santis}, \citenamefont
  {Loredo}, \citenamefont {Almeida}, \citenamefont {Hornecker}, \citenamefont
  {Portalupi}, \citenamefont {Grange}, \citenamefont {Anton}, \citenamefont
  {Demory} \emph {et~al.}}]{somaschi2016near}%
  \BibitemOpen
  \bibfield  {author} {\bibinfo {author} {\bibfnamefont {N.}~\bibnamefont
  {Somaschi}}, \bibinfo {author} {\bibfnamefont {V.}~\bibnamefont {Giesz}},
  \bibinfo {author} {\bibfnamefont {L.}~\bibnamefont {De~Santis}}, \bibinfo
  {author} {\bibfnamefont {J.}~\bibnamefont {Loredo}}, \bibinfo {author}
  {\bibfnamefont {M.~P.}\ \bibnamefont {Almeida}}, \bibinfo {author}
  {\bibfnamefont {G.}~\bibnamefont {Hornecker}}, \bibinfo {author}
  {\bibfnamefont {S.~L.}\ \bibnamefont {Portalupi}}, \bibinfo {author}
  {\bibfnamefont {T.}~\bibnamefont {Grange}}, \bibinfo {author} {\bibfnamefont
  {C.}~\bibnamefont {Anton}}, \bibinfo {author} {\bibfnamefont
  {J.}~\bibnamefont {Demory}},  \emph {et~al.},\ }\href
  {https://www.nature.com/articles/nphoton.2016.23} {\bibfield  {journal}
  {\bibinfo  {journal} {Nature Photonics}\ }\textbf {\bibinfo {volume} {10}},\
  \bibinfo {pages} {340} (\bibinfo {year} {2016})}\BibitemShut {NoStop}%
\bibitem [{\citenamefont {Loredo}\ \emph {et~al.}(2016)\citenamefont {Loredo},
  \citenamefont {Zakaria}, \citenamefont {Somaschi}, \citenamefont {Anton},
  \citenamefont {De~Santis}, \citenamefont {Giesz}, \citenamefont {Grange},
  \citenamefont {Broome}, \citenamefont {Gazzano}, \citenamefont {Coppola}
  \emph {et~al.}}]{loredo2016scalable}%
  \BibitemOpen
  \bibfield  {author} {\bibinfo {author} {\bibfnamefont {J.~C.}\ \bibnamefont
  {Loredo}}, \bibinfo {author} {\bibfnamefont {N.~A.}\ \bibnamefont {Zakaria}},
  \bibinfo {author} {\bibfnamefont {N.}~\bibnamefont {Somaschi}}, \bibinfo
  {author} {\bibfnamefont {C.}~\bibnamefont {Anton}}, \bibinfo {author}
  {\bibfnamefont {L.}~\bibnamefont {De~Santis}}, \bibinfo {author}
  {\bibfnamefont {V.}~\bibnamefont {Giesz}}, \bibinfo {author} {\bibfnamefont
  {T.}~\bibnamefont {Grange}}, \bibinfo {author} {\bibfnamefont {M.~A.}\
  \bibnamefont {Broome}}, \bibinfo {author} {\bibfnamefont {O.}~\bibnamefont
  {Gazzano}}, \bibinfo {author} {\bibfnamefont {G.}~\bibnamefont {Coppola}},
  \emph {et~al.},\ }\href
  {https://opg.optica.org/optica/fulltext.cfm?uri=optica-3-4-433&id=338998}
  {\bibfield  {journal} {\bibinfo  {journal} {Optica}\ }\textbf {\bibinfo
  {volume} {3}},\ \bibinfo {pages} {433} (\bibinfo {year} {2016})}\BibitemShut
  {NoStop}%
\bibitem [{\citenamefont {Gale}\ \emph {et~al.}(2021)\citenamefont {Gale},
  \citenamefont {Li}, \citenamefont {Chen}, \citenamefont {Watanabe},
  \citenamefont {Taniguchi}, \citenamefont {Aharonovich},\ and\ \citenamefont
  {Toth}}]{gale2021deterministic}%
  \BibitemOpen
  \bibfield  {author} {\bibinfo {author} {\bibfnamefont {A.}~\bibnamefont
  {Gale}}, \bibinfo {author} {\bibfnamefont {C.}~\bibnamefont {Li}}, \bibinfo
  {author} {\bibfnamefont {Y.}~\bibnamefont {Chen}}, \bibinfo {author}
  {\bibfnamefont {K.}~\bibnamefont {Watanabe}}, \bibinfo {author}
  {\bibfnamefont {T.}~\bibnamefont {Taniguchi}}, \bibinfo {author}
  {\bibfnamefont {I.}~\bibnamefont {Aharonovich}}, \ and\ \bibinfo {author}
  {\bibfnamefont {M.}~\bibnamefont {Toth}},\ }\href
  {https://arxiv.org/abs/2111.13441} {\bibfield  {journal} {\bibinfo  {journal}
  {arXiv preprint arXiv:2111.13441}\ } (\bibinfo {year} {2021})}\BibitemShut
  {NoStop}%
\bibitem [{\citenamefont {Koehl}\ \emph {et~al.}(2015)\citenamefont {Koehl},
  \citenamefont {Seo}, \citenamefont {Galli},\ and\ \citenamefont
  {Awschalom}}]{koehl2015designing}%
  \BibitemOpen
  \bibfield  {author} {\bibinfo {author} {\bibfnamefont {W.~F.}\ \bibnamefont
  {Koehl}}, \bibinfo {author} {\bibfnamefont {H.}~\bibnamefont {Seo}}, \bibinfo
  {author} {\bibfnamefont {G.}~\bibnamefont {Galli}}, \ and\ \bibinfo {author}
  {\bibfnamefont {D.~D.}\ \bibnamefont {Awschalom}},\ }\href
  {https://www.cambridge.org/core/journals/mrs-bulletin/article/designing-defect-spins-for-waferscale-quantum-technologies/5363ACC9111B063F946C214E94376ECA}
  {\bibfield  {journal} {\bibinfo  {journal} {MRS Bulletin}\ }\textbf {\bibinfo
  {volume} {40}},\ \bibinfo {pages} {1146} (\bibinfo {year}
  {2015})}\BibitemShut {NoStop}%
\bibitem [{\citenamefont {He}\ \emph {et~al.}(2015)\citenamefont {He},
  \citenamefont {Clark}, \citenamefont {Schaibley}, \citenamefont {He},
  \citenamefont {Chen}, \citenamefont {Wei}, \citenamefont {Ding},
  \citenamefont {Zhang}, \citenamefont {Yao}, \citenamefont {Xu} \emph
  {et~al.}}]{he2015single}%
  \BibitemOpen
  \bibfield  {author} {\bibinfo {author} {\bibfnamefont {Y.-M.}\ \bibnamefont
  {He}}, \bibinfo {author} {\bibfnamefont {G.}~\bibnamefont {Clark}}, \bibinfo
  {author} {\bibfnamefont {J.~R.}\ \bibnamefont {Schaibley}}, \bibinfo {author}
  {\bibfnamefont {Y.}~\bibnamefont {He}}, \bibinfo {author} {\bibfnamefont
  {M.-C.}\ \bibnamefont {Chen}}, \bibinfo {author} {\bibfnamefont {Y.-J.}\
  \bibnamefont {Wei}}, \bibinfo {author} {\bibfnamefont {X.}~\bibnamefont
  {Ding}}, \bibinfo {author} {\bibfnamefont {Q.}~\bibnamefont {Zhang}},
  \bibinfo {author} {\bibfnamefont {W.}~\bibnamefont {Yao}}, \bibinfo {author}
  {\bibfnamefont {X.}~\bibnamefont {Xu}},  \emph {et~al.},\ }\href
  {https://www.nature.com/articles/nnano.2015.75} {\bibfield  {journal}
  {\bibinfo  {journal} {Nature nanotechnology}\ }\textbf {\bibinfo {volume}
  {10}},\ \bibinfo {pages} {497} (\bibinfo {year} {2015})}\BibitemShut
  {NoStop}%
\bibitem [{\citenamefont {Toth}\ and\ \citenamefont
  {Aharonovich}(2019)}]{toth2019single}%
  \BibitemOpen
  \bibfield  {author} {\bibinfo {author} {\bibfnamefont {M.}~\bibnamefont
  {Toth}}\ and\ \bibinfo {author} {\bibfnamefont {I.}~\bibnamefont
  {Aharonovich}},\ }\href
  {https://www.annualreviews.org/doi/abs/10.1146/annurev-physchem-042018-052628}
  {\bibfield  {journal} {\bibinfo  {journal} {Annual review of physical
  chemistry}\ }\textbf {\bibinfo {volume} {70}},\ \bibinfo {pages} {123}
  (\bibinfo {year} {2019})}\BibitemShut {NoStop}%
\bibitem [{\citenamefont {Mendelson}\ \emph {et~al.}(2021)\citenamefont
  {Mendelson}, \citenamefont {Chugh}, \citenamefont {Reimers}, \citenamefont
  {Cheng}, \citenamefont {Gottscholl}, \citenamefont {Long}, \citenamefont
  {Mellor}, \citenamefont {Zettl}, \citenamefont {Dyakonov}, \citenamefont
  {Beton} \emph {et~al.}}]{mendelson2021identifying}%
  \BibitemOpen
  \bibfield  {author} {\bibinfo {author} {\bibfnamefont {N.}~\bibnamefont
  {Mendelson}}, \bibinfo {author} {\bibfnamefont {D.}~\bibnamefont {Chugh}},
  \bibinfo {author} {\bibfnamefont {J.~R.}\ \bibnamefont {Reimers}}, \bibinfo
  {author} {\bibfnamefont {T.~S.}\ \bibnamefont {Cheng}}, \bibinfo {author}
  {\bibfnamefont {A.}~\bibnamefont {Gottscholl}}, \bibinfo {author}
  {\bibfnamefont {H.}~\bibnamefont {Long}}, \bibinfo {author} {\bibfnamefont
  {C.~J.}\ \bibnamefont {Mellor}}, \bibinfo {author} {\bibfnamefont
  {A.}~\bibnamefont {Zettl}}, \bibinfo {author} {\bibfnamefont
  {V.}~\bibnamefont {Dyakonov}}, \bibinfo {author} {\bibfnamefont {P.~H.}\
  \bibnamefont {Beton}},  \emph {et~al.},\ }\href
  {https://www.nature.com/articles/s41563-020-00850-y} {\bibfield  {journal}
  {\bibinfo  {journal} {Nature materials}\ }\textbf {\bibinfo {volume} {20}},\
  \bibinfo {pages} {321} (\bibinfo {year} {2021})}\BibitemShut {NoStop}%
\bibitem [{\citenamefont {Tran}\ \emph {et~al.}(2016)\citenamefont {Tran},
  \citenamefont {Bray}, \citenamefont {Ford}, \citenamefont {Toth},\ and\
  \citenamefont {Aharonovich}}]{tran2016quantum}%
  \BibitemOpen
  \bibfield  {author} {\bibinfo {author} {\bibfnamefont {T.~T.}\ \bibnamefont
  {Tran}}, \bibinfo {author} {\bibfnamefont {K.}~\bibnamefont {Bray}}, \bibinfo
  {author} {\bibfnamefont {M.~J.}\ \bibnamefont {Ford}}, \bibinfo {author}
  {\bibfnamefont {M.}~\bibnamefont {Toth}}, \ and\ \bibinfo {author}
  {\bibfnamefont {I.}~\bibnamefont {Aharonovich}},\ }\href
  {https://www.nature.com/articles/nnano.2015.242} {\bibfield  {journal}
  {\bibinfo  {journal} {Nature Nanotechnology}\ }\textbf {\bibinfo {volume}
  {11}},\ \bibinfo {pages} {37} (\bibinfo {year} {2016})}\BibitemShut {NoStop}%
\bibitem [{\citenamefont {Li}\ \emph {et~al.}(2019)\citenamefont {Li},
  \citenamefont {Scully}, \citenamefont {Shayan}, \citenamefont {Luo},\ and\
  \citenamefont {Strauf}}]{li2019near}%
  \BibitemOpen
  \bibfield  {author} {\bibinfo {author} {\bibfnamefont {X.}~\bibnamefont
  {Li}}, \bibinfo {author} {\bibfnamefont {R.~A.}\ \bibnamefont {Scully}},
  \bibinfo {author} {\bibfnamefont {K.}~\bibnamefont {Shayan}}, \bibinfo
  {author} {\bibfnamefont {Y.}~\bibnamefont {Luo}}, \ and\ \bibinfo {author}
  {\bibfnamefont {S.}~\bibnamefont {Strauf}},\ }\href
  {https://pubs.acs.org/doi/abs/10.1021/acsnano.9b01996} {\bibfield  {journal}
  {\bibinfo  {journal} {ACS Nano}\ }\textbf {\bibinfo {volume} {13}},\ \bibinfo
  {pages} {6992} (\bibinfo {year} {2019})}\BibitemShut {NoStop}%
\bibitem [{\citenamefont {Kianinia}\ \emph {et~al.}(2017)\citenamefont
  {Kianinia}, \citenamefont {Regan}, \citenamefont {Tawfik}, \citenamefont
  {Tran}, \citenamefont {Ford}, \citenamefont {Aharonovich},\ and\
  \citenamefont {Toth}}]{kianinia2017robust}%
  \BibitemOpen
  \bibfield  {author} {\bibinfo {author} {\bibfnamefont {M.}~\bibnamefont
  {Kianinia}}, \bibinfo {author} {\bibfnamefont {B.}~\bibnamefont {Regan}},
  \bibinfo {author} {\bibfnamefont {S.~A.}\ \bibnamefont {Tawfik}}, \bibinfo
  {author} {\bibfnamefont {T.~T.}\ \bibnamefont {Tran}}, \bibinfo {author}
  {\bibfnamefont {M.~J.}\ \bibnamefont {Ford}}, \bibinfo {author}
  {\bibfnamefont {I.}~\bibnamefont {Aharonovich}}, \ and\ \bibinfo {author}
  {\bibfnamefont {M.}~\bibnamefont {Toth}},\ }\href
  {https://pubs.acs.org/doi/abs/10.1021/acsphotonics.7b00086} {\bibfield
  {journal} {\bibinfo  {journal} {ACS Photonics}\ }\textbf {\bibinfo {volume}
  {4}},\ \bibinfo {pages} {768} (\bibinfo {year} {2017})}\BibitemShut {NoStop}%
\bibitem [{\citenamefont {Proscia}\ \emph {et~al.}(2018)\citenamefont
  {Proscia}, \citenamefont {Shotan}, \citenamefont {Jayakumar}, \citenamefont
  {Reddy}, \citenamefont {Cohen}, \citenamefont {Dollar}, \citenamefont
  {Alkauskas}, \citenamefont {Doherty}, \citenamefont {Meriles},\ and\
  \citenamefont {Menon}}]{proscia2018near}%
  \BibitemOpen
  \bibfield  {author} {\bibinfo {author} {\bibfnamefont {N.~V.}\ \bibnamefont
  {Proscia}}, \bibinfo {author} {\bibfnamefont {Z.}~\bibnamefont {Shotan}},
  \bibinfo {author} {\bibfnamefont {H.}~\bibnamefont {Jayakumar}}, \bibinfo
  {author} {\bibfnamefont {P.}~\bibnamefont {Reddy}}, \bibinfo {author}
  {\bibfnamefont {C.}~\bibnamefont {Cohen}}, \bibinfo {author} {\bibfnamefont
  {M.}~\bibnamefont {Dollar}}, \bibinfo {author} {\bibfnamefont
  {A.}~\bibnamefont {Alkauskas}}, \bibinfo {author} {\bibfnamefont
  {M.}~\bibnamefont {Doherty}}, \bibinfo {author} {\bibfnamefont {C.~A.}\
  \bibnamefont {Meriles}}, \ and\ \bibinfo {author} {\bibfnamefont {V.~M.}\
  \bibnamefont {Menon}},\ }\href
  {https://opg.optica.org/optica/fulltext.cfm?uri=optica-5-9-1128&id=398241}
  {\bibfield  {journal} {\bibinfo  {journal} {Optica}\ }\textbf {\bibinfo
  {volume} {5}},\ \bibinfo {pages} {1128} (\bibinfo {year} {2018})}\BibitemShut
  {NoStop}%
\bibitem [{\citenamefont {Jungwirth}\ \emph {et~al.}(2016)\citenamefont
  {Jungwirth}, \citenamefont {Calderon}, \citenamefont {Ji}, \citenamefont
  {Spencer}, \citenamefont {Flatt{\'e}},\ and\ \citenamefont
  {Fuchs}}]{jungwirth2016temperature}%
  \BibitemOpen
  \bibfield  {author} {\bibinfo {author} {\bibfnamefont {N.~R.}\ \bibnamefont
  {Jungwirth}}, \bibinfo {author} {\bibfnamefont {B.}~\bibnamefont {Calderon}},
  \bibinfo {author} {\bibfnamefont {Y.}~\bibnamefont {Ji}}, \bibinfo {author}
  {\bibfnamefont {M.~G.}\ \bibnamefont {Spencer}}, \bibinfo {author}
  {\bibfnamefont {M.~E.}\ \bibnamefont {Flatt{\'e}}}, \ and\ \bibinfo {author}
  {\bibfnamefont {G.~D.}\ \bibnamefont {Fuchs}},\ }\href
  {https://pubs.acs.org/doi/abs/10.1021/acs.nanolett.6b01987} {\bibfield
  {journal} {\bibinfo  {journal} {Nano Letters}\ }\textbf {\bibinfo {volume}
  {16}},\ \bibinfo {pages} {6052} (\bibinfo {year} {2016})}\BibitemShut
  {NoStop}%
\bibitem [{\citenamefont {Bourrellier}\ \emph {et~al.}(2016)\citenamefont
  {Bourrellier}, \citenamefont {Meuret}, \citenamefont {Tararan}, \citenamefont
  {St{\'e}phan}, \citenamefont {Kociak}, \citenamefont {Tizei},\ and\
  \citenamefont {Zobelli}}]{bourrellier2016bright}%
  \BibitemOpen
  \bibfield  {author} {\bibinfo {author} {\bibfnamefont {R.}~\bibnamefont
  {Bourrellier}}, \bibinfo {author} {\bibfnamefont {S.}~\bibnamefont {Meuret}},
  \bibinfo {author} {\bibfnamefont {A.}~\bibnamefont {Tararan}}, \bibinfo
  {author} {\bibfnamefont {O.}~\bibnamefont {St{\'e}phan}}, \bibinfo {author}
  {\bibfnamefont {M.}~\bibnamefont {Kociak}}, \bibinfo {author} {\bibfnamefont
  {L.~H.}\ \bibnamefont {Tizei}}, \ and\ \bibinfo {author} {\bibfnamefont
  {A.}~\bibnamefont {Zobelli}},\ }\href
  {https://pubs.acs.org/doi/abs/10.1021/acs.nanolett.6b01368} {\bibfield
  {journal} {\bibinfo  {journal} {Nano Letters}\ }\textbf {\bibinfo {volume}
  {16}},\ \bibinfo {pages} {4317} (\bibinfo {year} {2016})}\BibitemShut
  {NoStop}%
\bibitem [{\citenamefont {Grosso}\ \emph {et~al.}(2017)\citenamefont {Grosso},
  \citenamefont {Moon}, \citenamefont {Lienhard}, \citenamefont {Ali},
  \citenamefont {Efetov}, \citenamefont {Furchi}, \citenamefont
  {Jarillo-Herrero}, \citenamefont {Ford}, \citenamefont {Aharonovich},\ and\
  \citenamefont {Englund}}]{grosso2017tunable}%
  \BibitemOpen
  \bibfield  {author} {\bibinfo {author} {\bibfnamefont {G.}~\bibnamefont
  {Grosso}}, \bibinfo {author} {\bibfnamefont {H.}~\bibnamefont {Moon}},
  \bibinfo {author} {\bibfnamefont {B.}~\bibnamefont {Lienhard}}, \bibinfo
  {author} {\bibfnamefont {S.}~\bibnamefont {Ali}}, \bibinfo {author}
  {\bibfnamefont {D.~K.}\ \bibnamefont {Efetov}}, \bibinfo {author}
  {\bibfnamefont {M.~M.}\ \bibnamefont {Furchi}}, \bibinfo {author}
  {\bibfnamefont {P.}~\bibnamefont {Jarillo-Herrero}}, \bibinfo {author}
  {\bibfnamefont {M.~J.}\ \bibnamefont {Ford}}, \bibinfo {author}
  {\bibfnamefont {I.}~\bibnamefont {Aharonovich}}, \ and\ \bibinfo {author}
  {\bibfnamefont {D.}~\bibnamefont {Englund}},\ }\href
  {https://www.nature.com/articles/s41467-017-00810-2} {\bibfield  {journal}
  {\bibinfo  {journal} {Nature Communications}\ }\textbf {\bibinfo {volume}
  {8}},\ \bibinfo {pages} {1} (\bibinfo {year} {2017})}\BibitemShut {NoStop}%
\bibitem [{\citenamefont {Zeng}\ \emph {et~al.}(2022)\citenamefont {Zeng},
  \citenamefont {Ngyuen}, \citenamefont {Ai}, \citenamefont {Bennet},
  \citenamefont {Solnstev}, \citenamefont {Laucht}, \citenamefont {Al-Juboori},
  \citenamefont {Toth}, \citenamefont {Mildren}, \citenamefont {Malaney} \emph
  {et~al.}}]{zeng2022integrated}%
  \BibitemOpen
  \bibfield  {author} {\bibinfo {author} {\bibfnamefont {H.~Z.~J.}\
  \bibnamefont {Zeng}}, \bibinfo {author} {\bibfnamefont {M.~A.~P.}\
  \bibnamefont {Ngyuen}}, \bibinfo {author} {\bibfnamefont {X.}~\bibnamefont
  {Ai}}, \bibinfo {author} {\bibfnamefont {A.}~\bibnamefont {Bennet}}, \bibinfo
  {author} {\bibfnamefont {A.~S.}\ \bibnamefont {Solnstev}}, \bibinfo {author}
  {\bibfnamefont {A.}~\bibnamefont {Laucht}}, \bibinfo {author} {\bibfnamefont
  {A.}~\bibnamefont {Al-Juboori}}, \bibinfo {author} {\bibfnamefont
  {M.}~\bibnamefont {Toth}}, \bibinfo {author} {\bibfnamefont {R.~P.}\
  \bibnamefont {Mildren}}, \bibinfo {author} {\bibfnamefont {R.}~\bibnamefont
  {Malaney}},  \emph {et~al.},\ }\href
  {https://opg.optica.org/ol/fulltext.cfm?uri=ol-47-7-1673&id=470703}
  {\bibfield  {journal} {\bibinfo  {journal} {Optics Letters}\ }\textbf
  {\bibinfo {volume} {47}},\ \bibinfo {pages} {1673} (\bibinfo {year}
  {2022})}\BibitemShut {NoStop}%
\bibitem [{\citenamefont {White}\ \emph {et~al.}(2021)\citenamefont {White},
  \citenamefont {Stewart}, \citenamefont {Solntsev}, \citenamefont {Li},
  \citenamefont {Toth}, \citenamefont {Kianinia},\ and\ \citenamefont
  {Aharonovich}}]{white2021phonon}%
  \BibitemOpen
  \bibfield  {author} {\bibinfo {author} {\bibfnamefont {S.}~\bibnamefont
  {White}}, \bibinfo {author} {\bibfnamefont {C.}~\bibnamefont {Stewart}},
  \bibinfo {author} {\bibfnamefont {A.~S.}\ \bibnamefont {Solntsev}}, \bibinfo
  {author} {\bibfnamefont {C.}~\bibnamefont {Li}}, \bibinfo {author}
  {\bibfnamefont {M.}~\bibnamefont {Toth}}, \bibinfo {author} {\bibfnamefont
  {M.}~\bibnamefont {Kianinia}}, \ and\ \bibinfo {author} {\bibfnamefont
  {I.}~\bibnamefont {Aharonovich}},\ }\href
  {https://opg.optica.org/optica/fulltext.cfm?uri=optica-8-9-1153&id=458165}
  {\bibfield  {journal} {\bibinfo  {journal} {Optica}\ }\textbf {\bibinfo
  {volume} {8}},\ \bibinfo {pages} {1153} (\bibinfo {year} {2021})}\BibitemShut
  {NoStop}%
\bibitem [{\citenamefont {H{\"o}gele}\ \emph {et~al.}(2004)\citenamefont
  {H{\"o}gele}, \citenamefont {Seidl}, \citenamefont {Kroner}, \citenamefont
  {Karrai}, \citenamefont {Warburton}, \citenamefont {Gerardot},\ and\
  \citenamefont {Petroff}}]{hogele2004voltage}%
  \BibitemOpen
  \bibfield  {author} {\bibinfo {author} {\bibfnamefont {A.}~\bibnamefont
  {H{\"o}gele}}, \bibinfo {author} {\bibfnamefont {S.}~\bibnamefont {Seidl}},
  \bibinfo {author} {\bibfnamefont {M.}~\bibnamefont {Kroner}}, \bibinfo
  {author} {\bibfnamefont {K.}~\bibnamefont {Karrai}}, \bibinfo {author}
  {\bibfnamefont {R.~J.}\ \bibnamefont {Warburton}}, \bibinfo {author}
  {\bibfnamefont {B.~D.}\ \bibnamefont {Gerardot}}, \ and\ \bibinfo {author}
  {\bibfnamefont {P.~M.}\ \bibnamefont {Petroff}},\ }\href
  {https://journals.aps.org/prl/abstract/10.1103/PhysRevLett.93.217401}
  {\bibfield  {journal} {\bibinfo  {journal} {Physical Review Letters}\
  }\textbf {\bibinfo {volume} {93}},\ \bibinfo {pages} {217401} (\bibinfo
  {year} {2004})}\BibitemShut {NoStop}%
\bibitem [{\citenamefont {Houel}\ \emph {et~al.}(2012)\citenamefont {Houel},
  \citenamefont {Kuhlmann}, \citenamefont {Greuter}, \citenamefont {Xue},
  \citenamefont {Poggio}, \citenamefont {Gerardot}, \citenamefont {Dalgarno},
  \citenamefont {Badolato}, \citenamefont {Petroff}, \citenamefont {Ludwig}
  \emph {et~al.}}]{houel2012probing}%
  \BibitemOpen
  \bibfield  {author} {\bibinfo {author} {\bibfnamefont {J.}~\bibnamefont
  {Houel}}, \bibinfo {author} {\bibfnamefont {A.}~\bibnamefont {Kuhlmann}},
  \bibinfo {author} {\bibfnamefont {L.}~\bibnamefont {Greuter}}, \bibinfo
  {author} {\bibfnamefont {F.}~\bibnamefont {Xue}}, \bibinfo {author}
  {\bibfnamefont {M.}~\bibnamefont {Poggio}}, \bibinfo {author} {\bibfnamefont
  {B.}~\bibnamefont {Gerardot}}, \bibinfo {author} {\bibfnamefont
  {P.}~\bibnamefont {Dalgarno}}, \bibinfo {author} {\bibfnamefont
  {A.}~\bibnamefont {Badolato}}, \bibinfo {author} {\bibfnamefont
  {P.}~\bibnamefont {Petroff}}, \bibinfo {author} {\bibfnamefont
  {A.}~\bibnamefont {Ludwig}},  \emph {et~al.},\ }\href
  {https://journals.aps.org/prl/abstract/10.1103/PhysRevLett.108.107401}
  {\bibfield  {journal} {\bibinfo  {journal} {Physical Review Letters}\
  }\textbf {\bibinfo {volume} {108}},\ \bibinfo {pages} {107401} (\bibinfo
  {year} {2012})}\BibitemShut {NoStop}%
\bibitem [{\citenamefont {Kuhlmann}\ \emph {et~al.}(2013)\citenamefont
  {Kuhlmann}, \citenamefont {Houel}, \citenamefont {Ludwig}, \citenamefont
  {Greuter}, \citenamefont {Reuter}, \citenamefont {Wieck}, \citenamefont
  {Poggio},\ and\ \citenamefont {Warburton}}]{kuhlmann2013charge}%
  \BibitemOpen
  \bibfield  {author} {\bibinfo {author} {\bibfnamefont {A.~V.}\ \bibnamefont
  {Kuhlmann}}, \bibinfo {author} {\bibfnamefont {J.}~\bibnamefont {Houel}},
  \bibinfo {author} {\bibfnamefont {A.}~\bibnamefont {Ludwig}}, \bibinfo
  {author} {\bibfnamefont {L.}~\bibnamefont {Greuter}}, \bibinfo {author}
  {\bibfnamefont {D.}~\bibnamefont {Reuter}}, \bibinfo {author} {\bibfnamefont
  {A.~D.}\ \bibnamefont {Wieck}}, \bibinfo {author} {\bibfnamefont
  {M.}~\bibnamefont {Poggio}}, \ and\ \bibinfo {author} {\bibfnamefont {R.~J.}\
  \bibnamefont {Warburton}},\ }\href
  {https://www.nature.com/articles/nphys2688} {\bibfield  {journal} {\bibinfo
  {journal} {Nature Physics}\ }\textbf {\bibinfo {volume} {9}},\ \bibinfo
  {pages} {570} (\bibinfo {year} {2013})}\BibitemShut {NoStop}%
\bibitem [{\citenamefont {Kuhlmann}\ \emph {et~al.}(2015)\citenamefont
  {Kuhlmann}, \citenamefont {Prechtel}, \citenamefont {Houel}, \citenamefont
  {Ludwig}, \citenamefont {Reuter}, \citenamefont {Wieck},\ and\ \citenamefont
  {Warburton}}]{kuhlmann2015transform}%
  \BibitemOpen
  \bibfield  {author} {\bibinfo {author} {\bibfnamefont {A.~V.}\ \bibnamefont
  {Kuhlmann}}, \bibinfo {author} {\bibfnamefont {J.~H.}\ \bibnamefont
  {Prechtel}}, \bibinfo {author} {\bibfnamefont {J.}~\bibnamefont {Houel}},
  \bibinfo {author} {\bibfnamefont {A.}~\bibnamefont {Ludwig}}, \bibinfo
  {author} {\bibfnamefont {D.}~\bibnamefont {Reuter}}, \bibinfo {author}
  {\bibfnamefont {A.~D.}\ \bibnamefont {Wieck}}, \ and\ \bibinfo {author}
  {\bibfnamefont {R.~J.}\ \bibnamefont {Warburton}},\ }\href
  {https://www.nature.com/articles/ncomms9204} {\bibfield  {journal} {\bibinfo
  {journal} {Nature Communications}\ }\textbf {\bibinfo {volume} {6}},\
  \bibinfo {pages} {1} (\bibinfo {year} {2015})}\BibitemShut {NoStop}%
\bibitem [{\citenamefont {Tamarat}\ \emph {et~al.}(2006)\citenamefont
  {Tamarat}, \citenamefont {Gaebel}, \citenamefont {Rabeau}, \citenamefont
  {Khan}, \citenamefont {Greentree}, \citenamefont {Wilson}, \citenamefont
  {Hollenberg}, \citenamefont {Prawer}, \citenamefont {Hemmer}, \citenamefont
  {Jelezko} \emph {et~al.}}]{tamarat2006stark}%
  \BibitemOpen
  \bibfield  {author} {\bibinfo {author} {\bibfnamefont {P.}~\bibnamefont
  {Tamarat}}, \bibinfo {author} {\bibfnamefont {T.}~\bibnamefont {Gaebel}},
  \bibinfo {author} {\bibfnamefont {J.}~\bibnamefont {Rabeau}}, \bibinfo
  {author} {\bibfnamefont {M.}~\bibnamefont {Khan}}, \bibinfo {author}
  {\bibfnamefont {A.}~\bibnamefont {Greentree}}, \bibinfo {author}
  {\bibfnamefont {H.}~\bibnamefont {Wilson}}, \bibinfo {author} {\bibfnamefont
  {L.}~\bibnamefont {Hollenberg}}, \bibinfo {author} {\bibfnamefont
  {S.}~\bibnamefont {Prawer}}, \bibinfo {author} {\bibfnamefont
  {P.}~\bibnamefont {Hemmer}}, \bibinfo {author} {\bibfnamefont
  {F.}~\bibnamefont {Jelezko}},  \emph {et~al.},\ }\href
  {https://journals.aps.org/prl/abstract/10.1103/PhysRevLett.97.083002}
  {\bibfield  {journal} {\bibinfo  {journal} {Physical Review Letters}\
  }\textbf {\bibinfo {volume} {97}},\ \bibinfo {pages} {083002} (\bibinfo
  {year} {2006})}\BibitemShut {NoStop}%
\bibitem [{\citenamefont {Batalov}\ \emph {et~al.}(2008)\citenamefont
  {Batalov}, \citenamefont {Zierl}, \citenamefont {Gaebel}, \citenamefont
  {Neumann}, \citenamefont {Chan}, \citenamefont {Balasubramanian},
  \citenamefont {Hemmer}, \citenamefont {Jelezko},\ and\ \citenamefont
  {Wrachtrup}}]{batalov2008temporal}%
  \BibitemOpen
  \bibfield  {author} {\bibinfo {author} {\bibfnamefont {A.}~\bibnamefont
  {Batalov}}, \bibinfo {author} {\bibfnamefont {C.}~\bibnamefont {Zierl}},
  \bibinfo {author} {\bibfnamefont {T.}~\bibnamefont {Gaebel}}, \bibinfo
  {author} {\bibfnamefont {P.}~\bibnamefont {Neumann}}, \bibinfo {author}
  {\bibfnamefont {I.-Y.}\ \bibnamefont {Chan}}, \bibinfo {author}
  {\bibfnamefont {G.}~\bibnamefont {Balasubramanian}}, \bibinfo {author}
  {\bibfnamefont {P.}~\bibnamefont {Hemmer}}, \bibinfo {author} {\bibfnamefont
  {F.}~\bibnamefont {Jelezko}}, \ and\ \bibinfo {author} {\bibfnamefont
  {J.}~\bibnamefont {Wrachtrup}},\ }\href
  {https://journals.aps.org/prl/abstract/10.1103/PhysRevLett.100.077401}
  {\bibfield  {journal} {\bibinfo  {journal} {Physical Review Letters}\
  }\textbf {\bibinfo {volume} {100}},\ \bibinfo {pages} {077401} (\bibinfo
  {year} {2008})}\BibitemShut {NoStop}%
\bibitem [{\citenamefont {Dietrich}\ \emph {et~al.}(2020)\citenamefont
  {Dietrich}, \citenamefont {Doherty}, \citenamefont {Aharonovich},\ and\
  \citenamefont {Kubanek}}]{dietrich2020solid}%
  \BibitemOpen
  \bibfield  {author} {\bibinfo {author} {\bibfnamefont {A.}~\bibnamefont
  {Dietrich}}, \bibinfo {author} {\bibfnamefont {M.}~\bibnamefont {Doherty}},
  \bibinfo {author} {\bibfnamefont {I.}~\bibnamefont {Aharonovich}}, \ and\
  \bibinfo {author} {\bibfnamefont {A.}~\bibnamefont {Kubanek}},\ }\href
  {https://journals.aps.org/prb/abstract/10.1103/PhysRevB.101.081401}
  {\bibfield  {journal} {\bibinfo  {journal} {Physical Review B}\ }\textbf
  {\bibinfo {volume} {101}},\ \bibinfo {pages} {081401} (\bibinfo {year}
  {2020})}\BibitemShut {NoStop}%
\bibitem [{\citenamefont {Hoese}\ \emph {et~al.}(2020)\citenamefont {Hoese},
  \citenamefont {Reddy}, \citenamefont {Dietrich}, \citenamefont {Koch},
  \citenamefont {Fehler}, \citenamefont {Doherty},\ and\ \citenamefont
  {Kubanek}}]{hoese2020mechanical}%
  \BibitemOpen
  \bibfield  {author} {\bibinfo {author} {\bibfnamefont {M.}~\bibnamefont
  {Hoese}}, \bibinfo {author} {\bibfnamefont {P.}~\bibnamefont {Reddy}},
  \bibinfo {author} {\bibfnamefont {A.}~\bibnamefont {Dietrich}}, \bibinfo
  {author} {\bibfnamefont {M.~K.}\ \bibnamefont {Koch}}, \bibinfo {author}
  {\bibfnamefont {K.~G.}\ \bibnamefont {Fehler}}, \bibinfo {author}
  {\bibfnamefont {M.~W.}\ \bibnamefont {Doherty}}, \ and\ \bibinfo {author}
  {\bibfnamefont {A.}~\bibnamefont {Kubanek}},\ }\href
  {https://www.science.org/doi/full/10.1126/sciadv.aba6038} {\bibfield
  {journal} {\bibinfo  {journal} {Science Advances}\ }\textbf {\bibinfo
  {volume} {6}},\ \bibinfo {pages} {eaba6038} (\bibinfo {year}
  {2020})}\BibitemShut {NoStop}%
\bibitem [{\citenamefont {Hepp}\ \emph {et~al.}(2014)\citenamefont {Hepp},
  \citenamefont {M{\"u}ller}, \citenamefont {Waselowski}, \citenamefont
  {Becker}, \citenamefont {Pingault}, \citenamefont {Sternschulte},
  \citenamefont {Steinm{\"u}ller-Nethl}, \citenamefont {Gali}, \citenamefont
  {Maze}, \citenamefont {Atat{\"u}re} \emph {et~al.}}]{hepp2014electronic}%
  \BibitemOpen
  \bibfield  {author} {\bibinfo {author} {\bibfnamefont {C.}~\bibnamefont
  {Hepp}}, \bibinfo {author} {\bibfnamefont {T.}~\bibnamefont {M{\"u}ller}},
  \bibinfo {author} {\bibfnamefont {V.}~\bibnamefont {Waselowski}}, \bibinfo
  {author} {\bibfnamefont {J.~N.}\ \bibnamefont {Becker}}, \bibinfo {author}
  {\bibfnamefont {B.}~\bibnamefont {Pingault}}, \bibinfo {author}
  {\bibfnamefont {H.}~\bibnamefont {Sternschulte}}, \bibinfo {author}
  {\bibfnamefont {D.}~\bibnamefont {Steinm{\"u}ller-Nethl}}, \bibinfo {author}
  {\bibfnamefont {A.}~\bibnamefont {Gali}}, \bibinfo {author} {\bibfnamefont
  {J.~R.}\ \bibnamefont {Maze}}, \bibinfo {author} {\bibfnamefont
  {M.}~\bibnamefont {Atat{\"u}re}},  \emph {et~al.},\ }\href
  {https://journals.aps.org/prl/abstract/10.1103/PhysRevLett.112.036405}
  {\bibfield  {journal} {\bibinfo  {journal} {Physical Review Letters}\
  }\textbf {\bibinfo {volume} {112}},\ \bibinfo {pages} {036405} (\bibinfo
  {year} {2014})}\BibitemShut {NoStop}%
\bibitem [{\citenamefont {Gali}\ \emph {et~al.}(2008)\citenamefont {Gali},
  \citenamefont {Fyta},\ and\ \citenamefont {Kaxiras}}]{gali2008ab}%
  \BibitemOpen
  \bibfield  {author} {\bibinfo {author} {\bibfnamefont {A.}~\bibnamefont
  {Gali}}, \bibinfo {author} {\bibfnamefont {M.}~\bibnamefont {Fyta}}, \ and\
  \bibinfo {author} {\bibfnamefont {E.}~\bibnamefont {Kaxiras}},\ }\href
  {https://journals.aps.org/prb/abstract/10.1103/PhysRevB.77.155206} {\bibfield
   {journal} {\bibinfo  {journal} {Physical Review B}\ }\textbf {\bibinfo
  {volume} {77}},\ \bibinfo {pages} {155206} (\bibinfo {year}
  {2008})}\BibitemShut {NoStop}%
\bibitem [{\citenamefont {Abdi}\ \emph {et~al.}(2018)\citenamefont {Abdi},
  \citenamefont {Chou}, \citenamefont {Gali},\ and\ \citenamefont
  {Plenio}}]{abdi2018color}%
  \BibitemOpen
  \bibfield  {author} {\bibinfo {author} {\bibfnamefont {M.}~\bibnamefont
  {Abdi}}, \bibinfo {author} {\bibfnamefont {J.-P.}\ \bibnamefont {Chou}},
  \bibinfo {author} {\bibfnamefont {A.}~\bibnamefont {Gali}}, \ and\ \bibinfo
  {author} {\bibfnamefont {M.~B.}\ \bibnamefont {Plenio}},\ }\href
  {https://pubs.acs.org/doi/abs/10.1021/acsphotonics.7b01442} {\bibfield
  {journal} {\bibinfo  {journal} {ACS Photonics}\ }\textbf {\bibinfo {volume}
  {5}},\ \bibinfo {pages} {1967} (\bibinfo {year} {2018})}\BibitemShut
  {NoStop}%
\bibitem [{\citenamefont {Ping}\ and\ \citenamefont
  {Smart}(2021)}]{ping2021computational}%
  \BibitemOpen
  \bibfield  {author} {\bibinfo {author} {\bibfnamefont {Y.}~\bibnamefont
  {Ping}}\ and\ \bibinfo {author} {\bibfnamefont {T.~J.}\ \bibnamefont
  {Smart}},\ }\href {https://www.nature.com/articles/s43588-021-00140-w}
  {\bibfield  {journal} {\bibinfo  {journal} {Nature Computational Science}\
  }\textbf {\bibinfo {volume} {1}},\ \bibinfo {pages} {646} (\bibinfo {year}
  {2021})}\BibitemShut {NoStop}%
\bibitem [{\citenamefont {Jara}\ \emph {et~al.}(2021)\citenamefont {Jara},
  \citenamefont {Rauch}, \citenamefont {Botti}, \citenamefont {Marques},
  \citenamefont {Norambuena}, \citenamefont {Coto}, \citenamefont
  {Castellanos-{\'A}guila}, \citenamefont {Maze},\ and\ \citenamefont
  {Munoz}}]{jara2021first}%
  \BibitemOpen
  \bibfield  {author} {\bibinfo {author} {\bibfnamefont {C.}~\bibnamefont
  {Jara}}, \bibinfo {author} {\bibfnamefont {T.}~\bibnamefont {Rauch}},
  \bibinfo {author} {\bibfnamefont {S.}~\bibnamefont {Botti}}, \bibinfo
  {author} {\bibfnamefont {M.~A.}\ \bibnamefont {Marques}}, \bibinfo {author}
  {\bibfnamefont {A.}~\bibnamefont {Norambuena}}, \bibinfo {author}
  {\bibfnamefont {R.}~\bibnamefont {Coto}}, \bibinfo {author} {\bibfnamefont
  {J.}~\bibnamefont {Castellanos-{\'A}guila}}, \bibinfo {author} {\bibfnamefont
  {J.~R.}\ \bibnamefont {Maze}}, \ and\ \bibinfo {author} {\bibfnamefont
  {F.}~\bibnamefont {Munoz}},\ }\href
  {https://pubs.acs.org/doi/abs/10.1021/acs.jpca.0c07339} {\bibfield  {journal}
  {\bibinfo  {journal} {The Journal of Physical Chemistry A}\ }\textbf
  {\bibinfo {volume} {125}},\ \bibinfo {pages} {1325} (\bibinfo {year}
  {2021})}\BibitemShut {NoStop}%
\bibitem [{\citenamefont {Bhang}\ \emph {et~al.}(2021)\citenamefont {Bhang},
  \citenamefont {Ma}, \citenamefont {Yim}, \citenamefont {Galli},\ and\
  \citenamefont {Seo}}]{bhang2021first}%
  \BibitemOpen
  \bibfield  {author} {\bibinfo {author} {\bibfnamefont {J.}~\bibnamefont
  {Bhang}}, \bibinfo {author} {\bibfnamefont {H.}~\bibnamefont {Ma}}, \bibinfo
  {author} {\bibfnamefont {D.}~\bibnamefont {Yim}}, \bibinfo {author}
  {\bibfnamefont {G.}~\bibnamefont {Galli}}, \ and\ \bibinfo {author}
  {\bibfnamefont {H.}~\bibnamefont {Seo}},\ }\href
  {https://pubs.acs.org/doi/full/10.1021/acsami.1c16988?casa_token=H-e8AdRdmIkAAAAA\%3Ah6quid6T5eAPJQPcNBgFvR4fSYT5dxoyRV4UJsJu8HzLjD6wPGoe7bkCvHBOXTe5UqbYh4mLVelrSxw}
  {\bibfield  {journal} {\bibinfo  {journal} {ACS Applied Materials \&
  Interfaces}\ }\textbf {\bibinfo {volume} {13}},\ \bibinfo {pages} {45768}
  (\bibinfo {year} {2021})}\BibitemShut {NoStop}%
\bibitem [{\citenamefont {Sajid}\ \emph {et~al.}(2018)\citenamefont {Sajid},
  \citenamefont {Reimers},\ and\ \citenamefont {Ford}}]{sajid2018defect}%
  \BibitemOpen
  \bibfield  {author} {\bibinfo {author} {\bibfnamefont {A.}~\bibnamefont
  {Sajid}}, \bibinfo {author} {\bibfnamefont {J.~R.}\ \bibnamefont {Reimers}},
  \ and\ \bibinfo {author} {\bibfnamefont {M.~J.}\ \bibnamefont {Ford}},\
  }\href {https://journals.aps.org/prb/abstract/10.1103/PhysRevB.97.064101}
  {\bibfield  {journal} {\bibinfo  {journal} {Physical Review B}\ }\textbf
  {\bibinfo {volume} {97}},\ \bibinfo {pages} {064101} (\bibinfo {year}
  {2018})}\BibitemShut {NoStop}%
\bibitem [{\citenamefont {Iv{\'a}dy}\ \emph {et~al.}(2020)\citenamefont
  {Iv{\'a}dy}, \citenamefont {Barcza}, \citenamefont {Thiering}, \citenamefont
  {Li}, \citenamefont {Hamdi}, \citenamefont {Chou}, \citenamefont {Legeza},\
  and\ \citenamefont {Gali}}]{ivady2020ab}%
  \BibitemOpen
  \bibfield  {author} {\bibinfo {author} {\bibfnamefont {V.}~\bibnamefont
  {Iv{\'a}dy}}, \bibinfo {author} {\bibfnamefont {G.}~\bibnamefont {Barcza}},
  \bibinfo {author} {\bibfnamefont {G.}~\bibnamefont {Thiering}}, \bibinfo
  {author} {\bibfnamefont {S.}~\bibnamefont {Li}}, \bibinfo {author}
  {\bibfnamefont {H.}~\bibnamefont {Hamdi}}, \bibinfo {author} {\bibfnamefont
  {J.-P.}\ \bibnamefont {Chou}}, \bibinfo {author} {\bibfnamefont
  {{\"O}.}~\bibnamefont {Legeza}}, \ and\ \bibinfo {author} {\bibfnamefont
  {A.}~\bibnamefont {Gali}},\ }\href
  {https://www.nature.com/articles/s41524-020-0305-x} {\bibfield  {journal}
  {\bibinfo  {journal} {npj Computational Materials}\ }\textbf {\bibinfo
  {volume} {6}},\ \bibinfo {pages} {1} (\bibinfo {year} {2020})}\BibitemShut
  {NoStop}%
\bibitem [{\citenamefont {Alkauskas}\ \emph {et~al.}(2014)\citenamefont
  {Alkauskas}, \citenamefont {Buckley}, \citenamefont {Awschalom},\ and\
  \citenamefont {Van~de Walle}}]{alkauskas2014first}%
  \BibitemOpen
  \bibfield  {author} {\bibinfo {author} {\bibfnamefont {A.}~\bibnamefont
  {Alkauskas}}, \bibinfo {author} {\bibfnamefont {B.~B.}\ \bibnamefont
  {Buckley}}, \bibinfo {author} {\bibfnamefont {D.~D.}\ \bibnamefont
  {Awschalom}}, \ and\ \bibinfo {author} {\bibfnamefont {C.~G.}\ \bibnamefont
  {Van~de Walle}},\ }\href
  {https://iopscience.iop.org/article/10.1088/1367-2630/16/7/073026/meta}
  {\bibfield  {journal} {\bibinfo  {journal} {New Journal of Physics}\ }\textbf
  {\bibinfo {volume} {16}},\ \bibinfo {pages} {073026} (\bibinfo {year}
  {2014})}\BibitemShut {NoStop}%
\bibitem [{\citenamefont {Razinkovas}\ \emph {et~al.}(2021)\citenamefont
  {Razinkovas}, \citenamefont {Doherty}, \citenamefont {Manson}, \citenamefont
  {Van~de Walle},\ and\ \citenamefont {Alkauskas}}]{razinkovas2021vibrational}%
  \BibitemOpen
  \bibfield  {author} {\bibinfo {author} {\bibfnamefont {L.}~\bibnamefont
  {Razinkovas}}, \bibinfo {author} {\bibfnamefont {M.~W.}\ \bibnamefont
  {Doherty}}, \bibinfo {author} {\bibfnamefont {N.~B.}\ \bibnamefont {Manson}},
  \bibinfo {author} {\bibfnamefont {C.~G.}\ \bibnamefont {Van~de Walle}}, \
  and\ \bibinfo {author} {\bibfnamefont {A.}~\bibnamefont {Alkauskas}},\ }\href
  {https://journals.aps.org/prb/abstract/10.1103/PhysRevB.104.045303}
  {\bibfield  {journal} {\bibinfo  {journal} {Physical Review B}\ }\textbf
  {\bibinfo {volume} {104}},\ \bibinfo {pages} {045303} (\bibinfo {year}
  {2021})}\BibitemShut {NoStop}%
\bibitem [{\citenamefont {Plakhotnik}\ \emph {et~al.}(2015)\citenamefont
  {Plakhotnik}, \citenamefont {Doherty},\ and\ \citenamefont
  {Manson}}]{plakhotnik2015electron}%
  \BibitemOpen
  \bibfield  {author} {\bibinfo {author} {\bibfnamefont {T.}~\bibnamefont
  {Plakhotnik}}, \bibinfo {author} {\bibfnamefont {M.~W.}\ \bibnamefont
  {Doherty}}, \ and\ \bibinfo {author} {\bibfnamefont {N.~B.}\ \bibnamefont
  {Manson}},\ }\href
  {https://journals.aps.org/prb/abstract/10.1103/PhysRevB.92.081203} {\bibfield
   {journal} {\bibinfo  {journal} {Physical Review B}\ }\textbf {\bibinfo
  {volume} {92}},\ \bibinfo {pages} {081203} (\bibinfo {year}
  {2015})}\BibitemShut {NoStop}%
\bibitem [{\citenamefont {Kohn}\ and\ \citenamefont
  {Sham}(1965)}]{kohn1965self}%
  \BibitemOpen
  \bibfield  {author} {\bibinfo {author} {\bibfnamefont {W.}~\bibnamefont
  {Kohn}}\ and\ \bibinfo {author} {\bibfnamefont {L.~J.}\ \bibnamefont
  {Sham}},\ }\href
  {https://journals.aps.org/pr/abstract/10.1103/PhysRev.140.A1133} {\bibfield
  {journal} {\bibinfo  {journal} {Physical Review}\ }\textbf {\bibinfo {volume}
  {140}},\ \bibinfo {pages} {A1133} (\bibinfo {year} {1965})}\BibitemShut
  {NoStop}%
\bibitem [{\citenamefont {Giustino}(2017)}]{giustino2017electron}%
  \BibitemOpen
  \bibfield  {author} {\bibinfo {author} {\bibfnamefont {F.}~\bibnamefont
  {Giustino}},\ }\href
  {https://journals.aps.org/rmp/abstract/10.1103/RevModPhys.89.015003}
  {\bibfield  {journal} {\bibinfo  {journal} {Reviews of Modern Physics}\
  }\textbf {\bibinfo {volume} {89}},\ \bibinfo {pages} {015003} (\bibinfo
  {year} {2017})}\BibitemShut {NoStop}%
\bibitem [{\citenamefont {Maradudin}(1966)}]{maradudin1966solid}%
  \BibitemOpen
  \bibfield  {author} {\bibinfo {author} {\bibfnamefont {A.~A.}\ \bibnamefont
  {Maradudin}},\ }\href@noop {} {\emph {\bibinfo {title} {Solid State
  Physics}}},\ edited by\ \bibinfo {editor} {\bibfnamefont {F.}~\bibnamefont
  {Seitz}}\ and\ \bibinfo {editor} {\bibfnamefont {D.}~\bibnamefont
  {Turnbull}},\ Vol.~\bibinfo {volume} {18}\ (\bibinfo  {publisher} {Academic
  Press, New York},\ \bibinfo {year} {1966})\ pp.\ \bibinfo {pages}
  {273--420}\BibitemShut {NoStop}%
\bibitem [{\citenamefont {Davies}(1974)}]{davies1974vibronic}%
  \BibitemOpen
  \bibfield  {author} {\bibinfo {author} {\bibfnamefont {G.}~\bibnamefont
  {Davies}},\ }\href
  {https://iopscience.iop.org/article/10.1088/0022-3719/7/20/019/meta}
  {\bibfield  {journal} {\bibinfo  {journal} {Journal of Physics C: Solid State
  Physics}\ }\textbf {\bibinfo {volume} {7}},\ \bibinfo {pages} {3797}
  (\bibinfo {year} {1974})}\BibitemShut {NoStop}%
\bibitem [{\citenamefont {Fu}\ \emph {et~al.}(2009)\citenamefont {Fu},
  \citenamefont {Santori}, \citenamefont {Barclay}, \citenamefont {Rogers},
  \citenamefont {Manson},\ and\ \citenamefont
  {Beausoleil}}]{fu2009observation}%
  \BibitemOpen
  \bibfield  {author} {\bibinfo {author} {\bibfnamefont {K.-M.~C.}\
  \bibnamefont {Fu}}, \bibinfo {author} {\bibfnamefont {C.}~\bibnamefont
  {Santori}}, \bibinfo {author} {\bibfnamefont {P.~E.}\ \bibnamefont
  {Barclay}}, \bibinfo {author} {\bibfnamefont {L.~J.}\ \bibnamefont {Rogers}},
  \bibinfo {author} {\bibfnamefont {N.~B.}\ \bibnamefont {Manson}}, \ and\
  \bibinfo {author} {\bibfnamefont {R.~G.}\ \bibnamefont {Beausoleil}},\ }\href
  {https://journals.aps.org/prl/abstract/10.1103/PhysRevLett.103.256404}
  {\bibfield  {journal} {\bibinfo  {journal} {Physical Review Letters}\
  }\textbf {\bibinfo {volume} {103}},\ \bibinfo {pages} {256404} (\bibinfo
  {year} {2009})}\BibitemShut {NoStop}%
\bibitem [{\citenamefont {Abtew}\ \emph {et~al.}(2011)\citenamefont {Abtew},
  \citenamefont {Sun}, \citenamefont {Shih}, \citenamefont {Dev}, \citenamefont
  {Zhang},\ and\ \citenamefont {Zhang}}]{abtew2011dynamic}%
  \BibitemOpen
  \bibfield  {author} {\bibinfo {author} {\bibfnamefont {T.~A.}\ \bibnamefont
  {Abtew}}, \bibinfo {author} {\bibfnamefont {Y.}~\bibnamefont {Sun}}, \bibinfo
  {author} {\bibfnamefont {B.-C.}\ \bibnamefont {Shih}}, \bibinfo {author}
  {\bibfnamefont {P.}~\bibnamefont {Dev}}, \bibinfo {author} {\bibfnamefont
  {S.}~\bibnamefont {Zhang}}, \ and\ \bibinfo {author} {\bibfnamefont
  {P.}~\bibnamefont {Zhang}},\ }\href
  {https://journals.aps.org/prl/abstract/10.1103/PhysRevLett.107.146403}
  {\bibfield  {journal} {\bibinfo  {journal} {Physical Review Letters}\
  }\textbf {\bibinfo {volume} {107}},\ \bibinfo {pages} {146403} (\bibinfo
  {year} {2011})}\BibitemShut {NoStop}%
\bibitem [{\citenamefont {Jahnke}\ \emph {et~al.}(2015)\citenamefont {Jahnke},
  \citenamefont {Sipahigil}, \citenamefont {Binder}, \citenamefont {Doherty},
  \citenamefont {Metsch}, \citenamefont {Rogers}, \citenamefont {Manson},
  \citenamefont {Lukin},\ and\ \citenamefont {Jelezko}}]{jahnke2015electron}%
  \BibitemOpen
  \bibfield  {author} {\bibinfo {author} {\bibfnamefont {K.~D.}\ \bibnamefont
  {Jahnke}}, \bibinfo {author} {\bibfnamefont {A.}~\bibnamefont {Sipahigil}},
  \bibinfo {author} {\bibfnamefont {J.~M.}\ \bibnamefont {Binder}}, \bibinfo
  {author} {\bibfnamefont {M.~W.}\ \bibnamefont {Doherty}}, \bibinfo {author}
  {\bibfnamefont {M.}~\bibnamefont {Metsch}}, \bibinfo {author} {\bibfnamefont
  {L.~J.}\ \bibnamefont {Rogers}}, \bibinfo {author} {\bibfnamefont {N.~B.}\
  \bibnamefont {Manson}}, \bibinfo {author} {\bibfnamefont {M.~D.}\
  \bibnamefont {Lukin}}, \ and\ \bibinfo {author} {\bibfnamefont
  {F.}~\bibnamefont {Jelezko}},\ }\href
  {https://iopscience.iop.org/article/10.1088/1367-2630/17/4/043011/meta}
  {\bibfield  {journal} {\bibinfo  {journal} {New Journal of Physics}\ }\textbf
  {\bibinfo {volume} {17}},\ \bibinfo {pages} {043011} (\bibinfo {year}
  {2015})}\BibitemShut {NoStop}%
\bibitem [{\citenamefont {Sakurai}\ and\ \citenamefont
  {Napolitano}(2021)}]{sakurai2021modern}%
  \BibitemOpen
  \bibfield  {author} {\bibinfo {author} {\bibfnamefont {J.}~\bibnamefont
  {Sakurai}}\ and\ \bibinfo {author} {\bibfnamefont {J.}~\bibnamefont
  {Napolitano}},\ }\href@noop {} {\emph {\bibinfo {title} {Modern Quantum
  Mechanics 3rd ed.}}}\ (\bibinfo  {publisher} {Cambridge University Press},\
  \bibinfo {address} {Cambridge, England},\ \bibinfo {year} {2021})\BibitemShut
  {NoStop}%
\bibitem [{\citenamefont {Golami}\ \emph {et~al.}(2022)\citenamefont {Golami},
  \citenamefont {Sharman}, \citenamefont {Ghobadi}, \citenamefont {Wein},
  \citenamefont {Zadeh-Haghighi}, \citenamefont {da~Rocha}, \citenamefont
  {Salahub},\ and\ \citenamefont {Simon}}]{golami2022b}%
  \BibitemOpen
  \bibfield  {author} {\bibinfo {author} {\bibfnamefont {O.}~\bibnamefont
  {Golami}}, \bibinfo {author} {\bibfnamefont {K.}~\bibnamefont {Sharman}},
  \bibinfo {author} {\bibfnamefont {R.}~\bibnamefont {Ghobadi}}, \bibinfo
  {author} {\bibfnamefont {S.~C.}\ \bibnamefont {Wein}}, \bibinfo {author}
  {\bibfnamefont {H.}~\bibnamefont {Zadeh-Haghighi}}, \bibinfo {author}
  {\bibfnamefont {C.~G.}\ \bibnamefont {da~Rocha}}, \bibinfo {author}
  {\bibfnamefont {D.~R.}\ \bibnamefont {Salahub}}, \ and\ \bibinfo {author}
  {\bibfnamefont {C.}~\bibnamefont {Simon}},\ }\href
  {https://journals.aps.org/prb/abstract/10.1103/PhysRevB.105.184101}
  {\bibfield  {journal} {\bibinfo  {journal} {Physical Review B}\ }\textbf
  {\bibinfo {volume} {105}},\ \bibinfo {pages} {184101} (\bibinfo {year}
  {2022})}\BibitemShut {NoStop}%
\bibitem [{\citenamefont {Heid}(2013)}]{heid201312}%
  \BibitemOpen
  \bibfield  {author} {\bibinfo {author} {\bibfnamefont {R.}~\bibnamefont
  {Heid}},\ }\href
  {https://books.google.ca/books?hl=en&lr=&id=YZSz2n6DiFUC&oi=fnd&pg=SA12-PA1&dq=Density+Functional+Perturbation+Theory+and+Electron+Phonon+Coupling&ots=wKsDOtHbYC&sig=10bjISEce_3w6ygReJa0KsdPTi0#v=onepage&q=Density\%20Functional\%20Perturbation\%20Theory\%20and\%20Electron\%20Phonon\%20Coupling&f=false}
  {\bibfield  {journal} {\bibinfo  {journal} {Emergent Phenomena in Correlated
  Matter}\ } (\bibinfo {year} {2013})}\BibitemShut {NoStop}%
\bibitem [{\citenamefont {Verstraete}\ and\ \citenamefont
  {Zanolli}(2014)}]{verstraete2014c}%
  \BibitemOpen
  \bibfield  {author} {\bibinfo {author} {\bibfnamefont {M.}~\bibnamefont
  {Verstraete}}\ and\ \bibinfo {author} {\bibfnamefont {Z.}~\bibnamefont
  {Zanolli}},\ }\href
  {https://www.researchgate.net/profile/Zeila-Zanolli/publication/283014374_Density_Functional_Perturbation_Theory/links/56264d9508aed3d3f13827ce/Density-Functional-Perturbation-Theory.pdf}
  {\  (\bibinfo {year} {2014})}\BibitemShut {NoStop}%
\bibitem [{\citenamefont {Srivastava}(2019)}]{srivastava2019physics}%
  \BibitemOpen
  \bibfield  {author} {\bibinfo {author} {\bibfnamefont {G.~P.}\ \bibnamefont
  {Srivastava}},\ }\href
  {https://www.taylorfrancis.com/books/mono/10.1201/9780203736241/physics-phonons-srivastava}
  {\emph {\bibinfo {title} {The physics of phonons}}}\ (\bibinfo  {publisher}
  {Routledge},\ \bibinfo {year} {2019})\BibitemShut {NoStop}%
\bibitem [{\citenamefont {Togo}\ and\ \citenamefont
  {Tanaka}(2015)}]{togo2015first}%
  \BibitemOpen
  \bibfield  {author} {\bibinfo {author} {\bibfnamefont {A.}~\bibnamefont
  {Togo}}\ and\ \bibinfo {author} {\bibfnamefont {I.}~\bibnamefont {Tanaka}},\
  }\href {https://www.sciencedirect.com/science/article/pii/S1359646215003127}
  {\bibfield  {journal} {\bibinfo  {journal} {Scripta Materialia}\ }\textbf
  {\bibinfo {volume} {108}},\ \bibinfo {pages} {1} (\bibinfo {year}
  {2015})}\BibitemShut {NoStop}%
\bibitem [{\citenamefont {Giustino}\ \emph {et~al.}(2007)\citenamefont
  {Giustino}, \citenamefont {Cohen},\ and\ \citenamefont
  {Louie}}]{giustino2007electron}%
  \BibitemOpen
  \bibfield  {author} {\bibinfo {author} {\bibfnamefont {F.}~\bibnamefont
  {Giustino}}, \bibinfo {author} {\bibfnamefont {M.~L.}\ \bibnamefont {Cohen}},
  \ and\ \bibinfo {author} {\bibfnamefont {S.~G.}\ \bibnamefont {Louie}},\
  }\href {https://journals.aps.org/prb/abstract/10.1103/PhysRevB.76.165108}
  {\bibfield  {journal} {\bibinfo  {journal} {Physical Review B}\ }\textbf
  {\bibinfo {volume} {76}},\ \bibinfo {pages} {165108} (\bibinfo {year}
  {2007})}\BibitemShut {NoStop}%
\bibitem [{\citenamefont {Ponc{\'e}}\ \emph {et~al.}(2016)\citenamefont
  {Ponc{\'e}}, \citenamefont {Margine}, \citenamefont {Verdi},\ and\
  \citenamefont {Giustino}}]{ponce2016epw}%
  \BibitemOpen
  \bibfield  {author} {\bibinfo {author} {\bibfnamefont {S.}~\bibnamefont
  {Ponc{\'e}}}, \bibinfo {author} {\bibfnamefont {E.~R.}\ \bibnamefont
  {Margine}}, \bibinfo {author} {\bibfnamefont {C.}~\bibnamefont {Verdi}}, \
  and\ \bibinfo {author} {\bibfnamefont {F.}~\bibnamefont {Giustino}},\ }\href
  {https://www.sciencedirect.com/science/article/pii/S0010465516302260}
  {\bibfield  {journal} {\bibinfo  {journal} {Computer Physics Communications}\
  }\textbf {\bibinfo {volume} {209}},\ \bibinfo {pages} {116} (\bibinfo {year}
  {2016})}\BibitemShut {NoStop}%
\bibitem [{\citenamefont {Li}\ \emph {et~al.}(2020)\citenamefont {Li},
  \citenamefont {Chou}, \citenamefont {Hu}, \citenamefont {Plenio},
  \citenamefont {Udvarhelyi}, \citenamefont {Thiering}, \citenamefont {Abdi},\
  and\ \citenamefont {Gali}}]{li2020giant}%
  \BibitemOpen
  \bibfield  {author} {\bibinfo {author} {\bibfnamefont {S.}~\bibnamefont
  {Li}}, \bibinfo {author} {\bibfnamefont {J.-P.}\ \bibnamefont {Chou}},
  \bibinfo {author} {\bibfnamefont {A.}~\bibnamefont {Hu}}, \bibinfo {author}
  {\bibfnamefont {M.~B.}\ \bibnamefont {Plenio}}, \bibinfo {author}
  {\bibfnamefont {P.}~\bibnamefont {Udvarhelyi}}, \bibinfo {author}
  {\bibfnamefont {G.}~\bibnamefont {Thiering}}, \bibinfo {author}
  {\bibfnamefont {M.}~\bibnamefont {Abdi}}, \ and\ \bibinfo {author}
  {\bibfnamefont {A.}~\bibnamefont {Gali}},\ }\href
  {https://www.nature.com/articles/s41534-020-00312-y} {\bibfield  {journal}
  {\bibinfo  {journal} {npj Quantum Information}\ }\textbf {\bibinfo {volume}
  {6}},\ \bibinfo {pages} {1} (\bibinfo {year} {2020})}\BibitemShut {NoStop}%
\bibitem [{\citenamefont {Noh}\ \emph {et~al.}(2018)\citenamefont {Noh},
  \citenamefont {Choi}, \citenamefont {Kim}, \citenamefont {Im}, \citenamefont
  {Kim}, \citenamefont {Seo},\ and\ \citenamefont {Lee}}]{noh2018stark}%
  \BibitemOpen
  \bibfield  {author} {\bibinfo {author} {\bibfnamefont {G.}~\bibnamefont
  {Noh}}, \bibinfo {author} {\bibfnamefont {D.}~\bibnamefont {Choi}}, \bibinfo
  {author} {\bibfnamefont {J.-H.}\ \bibnamefont {Kim}}, \bibinfo {author}
  {\bibfnamefont {D.-G.}\ \bibnamefont {Im}}, \bibinfo {author} {\bibfnamefont
  {Y.-H.}\ \bibnamefont {Kim}}, \bibinfo {author} {\bibfnamefont
  {H.}~\bibnamefont {Seo}}, \ and\ \bibinfo {author} {\bibfnamefont
  {J.}~\bibnamefont {Lee}},\ }\href
  {https://pubs.acs.org/doi/full/10.1021/acs.nanolett.8b01030} {\bibfield
  {journal} {\bibinfo  {journal} {Nano Letters}\ }\textbf {\bibinfo {volume}
  {18}},\ \bibinfo {pages} {4710} (\bibinfo {year} {2018})}\BibitemShut
  {NoStop}%
\bibitem [{\citenamefont {Gao}\ \emph {et~al.}(2021)\citenamefont {Gao},
  \citenamefont {Chen},\ and\ \citenamefont {Bernardi}}]{gao2021radiative}%
  \BibitemOpen
  \bibfield  {author} {\bibinfo {author} {\bibfnamefont {S.}~\bibnamefont
  {Gao}}, \bibinfo {author} {\bibfnamefont {H.-Y.}\ \bibnamefont {Chen}}, \
  and\ \bibinfo {author} {\bibfnamefont {M.}~\bibnamefont {Bernardi}},\ }\href
  {https://www.nature.com/articles/s41524-021-00544-2} {\bibfield  {journal}
  {\bibinfo  {journal} {npj Computational Materials}\ }\textbf {\bibinfo
  {volume} {7}},\ \bibinfo {pages} {1} (\bibinfo {year} {2021})}\BibitemShut
  {NoStop}%
\bibitem [{\citenamefont {Giannozzi}\ \emph {et~al.}(2009)\citenamefont
  {Giannozzi}, \citenamefont {Baroni}, \citenamefont {Bonini}, \citenamefont
  {Calandra}, \citenamefont {Car}, \citenamefont {Cavazzoni}, \citenamefont
  {Ceresoli}, \citenamefont {Chiarotti}, \citenamefont {Cococcioni},
  \citenamefont {Dabo} \emph {et~al.}}]{giannozzi2009quantum}%
  \BibitemOpen
  \bibfield  {author} {\bibinfo {author} {\bibfnamefont {P.}~\bibnamefont
  {Giannozzi}}, \bibinfo {author} {\bibfnamefont {S.}~\bibnamefont {Baroni}},
  \bibinfo {author} {\bibfnamefont {N.}~\bibnamefont {Bonini}}, \bibinfo
  {author} {\bibfnamefont {M.}~\bibnamefont {Calandra}}, \bibinfo {author}
  {\bibfnamefont {R.}~\bibnamefont {Car}}, \bibinfo {author} {\bibfnamefont
  {C.}~\bibnamefont {Cavazzoni}}, \bibinfo {author} {\bibfnamefont
  {D.}~\bibnamefont {Ceresoli}}, \bibinfo {author} {\bibfnamefont {G.~L.}\
  \bibnamefont {Chiarotti}}, \bibinfo {author} {\bibfnamefont {M.}~\bibnamefont
  {Cococcioni}}, \bibinfo {author} {\bibfnamefont {I.}~\bibnamefont {Dabo}},
  \emph {et~al.},\ }\href
  {https://iopscience.iop.org/article/10.1088/0953-8984/21/39/395502/meta}
  {\bibfield  {journal} {\bibinfo  {journal} {Journal of Physics: Condensed
  Matter}\ }\textbf {\bibinfo {volume} {21}},\ \bibinfo {pages} {395502}
  (\bibinfo {year} {2009})}\BibitemShut {NoStop}%
\bibitem [{\citenamefont {Giannozzi}\ \emph {et~al.}(2017)\citenamefont
  {Giannozzi}, \citenamefont {Andreussi}, \citenamefont {Brumme}, \citenamefont
  {Bunau}, \citenamefont {Nardelli}, \citenamefont {Calandra}, \citenamefont
  {Car}, \citenamefont {Cavazzoni}, \citenamefont {Ceresoli}, \citenamefont
  {Cococcioni} \emph {et~al.}}]{giannozzi2017advanced}%
  \BibitemOpen
  \bibfield  {author} {\bibinfo {author} {\bibfnamefont {P.}~\bibnamefont
  {Giannozzi}}, \bibinfo {author} {\bibfnamefont {O.}~\bibnamefont
  {Andreussi}}, \bibinfo {author} {\bibfnamefont {T.}~\bibnamefont {Brumme}},
  \bibinfo {author} {\bibfnamefont {O.}~\bibnamefont {Bunau}}, \bibinfo
  {author} {\bibfnamefont {M.~B.}\ \bibnamefont {Nardelli}}, \bibinfo {author}
  {\bibfnamefont {M.}~\bibnamefont {Calandra}}, \bibinfo {author}
  {\bibfnamefont {R.}~\bibnamefont {Car}}, \bibinfo {author} {\bibfnamefont
  {C.}~\bibnamefont {Cavazzoni}}, \bibinfo {author} {\bibfnamefont
  {D.}~\bibnamefont {Ceresoli}}, \bibinfo {author} {\bibfnamefont
  {M.}~\bibnamefont {Cococcioni}},  \emph {et~al.},\ }\href
  {https://iopscience.iop.org/article/10.1088/1361-648X/aa8f79/meta} {\bibfield
   {journal} {\bibinfo  {journal} {Journal of Physics: Condensed Matter}\
  }\textbf {\bibinfo {volume} {29}},\ \bibinfo {pages} {465901} (\bibinfo
  {year} {2017})}\BibitemShut {NoStop}%
\bibitem [{\citenamefont {Giannozzi}\ \emph {et~al.}(2020)\citenamefont
  {Giannozzi}, \citenamefont {Baseggio}, \citenamefont {Bonf{\`a}},
  \citenamefont {Brunato}, \citenamefont {Car}, \citenamefont {Carnimeo},
  \citenamefont {Cavazzoni}, \citenamefont {De~Gironcoli}, \citenamefont
  {Delugas}, \citenamefont {Ferrari~Ruffino} \emph
  {et~al.}}]{giannozzi2020quantum}%
  \BibitemOpen
  \bibfield  {author} {\bibinfo {author} {\bibfnamefont {P.}~\bibnamefont
  {Giannozzi}}, \bibinfo {author} {\bibfnamefont {O.}~\bibnamefont {Baseggio}},
  \bibinfo {author} {\bibfnamefont {P.}~\bibnamefont {Bonf{\`a}}}, \bibinfo
  {author} {\bibfnamefont {D.}~\bibnamefont {Brunato}}, \bibinfo {author}
  {\bibfnamefont {R.}~\bibnamefont {Car}}, \bibinfo {author} {\bibfnamefont
  {I.}~\bibnamefont {Carnimeo}}, \bibinfo {author} {\bibfnamefont
  {C.}~\bibnamefont {Cavazzoni}}, \bibinfo {author} {\bibfnamefont
  {S.}~\bibnamefont {De~Gironcoli}}, \bibinfo {author} {\bibfnamefont
  {P.}~\bibnamefont {Delugas}}, \bibinfo {author} {\bibfnamefont
  {F.}~\bibnamefont {Ferrari~Ruffino}},  \emph {et~al.},\ }\href
  {https://aip.scitation.org/doi/full/10.1063/5.0005082} {\bibfield  {journal}
  {\bibinfo  {journal} {The Journal of Chemical Physics}\ }\textbf {\bibinfo
  {volume} {152}},\ \bibinfo {pages} {154105} (\bibinfo {year}
  {2020})}\BibitemShut {NoStop}%
\bibitem [{\citenamefont {Noffsinger}\ \emph {et~al.}(2010)\citenamefont
  {Noffsinger}, \citenamefont {Giustino}, \citenamefont {Malone}, \citenamefont
  {Park}, \citenamefont {Louie},\ and\ \citenamefont
  {Cohen}}]{noffsinger2010epw}%
  \BibitemOpen
  \bibfield  {author} {\bibinfo {author} {\bibfnamefont {J.}~\bibnamefont
  {Noffsinger}}, \bibinfo {author} {\bibfnamefont {F.}~\bibnamefont
  {Giustino}}, \bibinfo {author} {\bibfnamefont {B.~D.}\ \bibnamefont
  {Malone}}, \bibinfo {author} {\bibfnamefont {C.-H.}\ \bibnamefont {Park}},
  \bibinfo {author} {\bibfnamefont {S.~G.}\ \bibnamefont {Louie}}, \ and\
  \bibinfo {author} {\bibfnamefont {M.~L.}\ \bibnamefont {Cohen}},\ }\href
  {https://www.sciencedirect.com/science/article/pii/S0010465510003218?casa_token=4jNpvEGuTTwAAAAA:jrHpKDVZyiF-vqtSQhEeX_r-5URHZ_BX0KXBS0YxNI2NzZiZxJ9MTxNyKXBN8_WO6T3DZeQPppIz}
  {\bibfield  {journal} {\bibinfo  {journal} {Computer Physics Communications}\
  }\textbf {\bibinfo {volume} {181}},\ \bibinfo {pages} {2140} (\bibinfo {year}
  {2010})}\BibitemShut {NoStop}%
\bibitem [{\citenamefont {Perdew}\ \emph {et~al.}(1996)\citenamefont {Perdew},
  \citenamefont {Burke},\ and\ \citenamefont
  {Ernzerhof}}]{perdew1996generalized}%
  \BibitemOpen
  \bibfield  {author} {\bibinfo {author} {\bibfnamefont {J.~P.}\ \bibnamefont
  {Perdew}}, \bibinfo {author} {\bibfnamefont {K.}~\bibnamefont {Burke}}, \
  and\ \bibinfo {author} {\bibfnamefont {M.}~\bibnamefont {Ernzerhof}},\ }\href
  {https://journals.aps.org/prl/abstract/10.1103/PhysRevLett.77.3865}
  {\bibfield  {journal} {\bibinfo  {journal} {Physical Review Letters}\
  }\textbf {\bibinfo {volume} {77}},\ \bibinfo {pages} {3865} (\bibinfo {year}
  {1996})}\BibitemShut {NoStop}%
\bibitem [{\citenamefont {Bl{\"o}chl}(1994)}]{blochl1994projector}%
  \BibitemOpen
  \bibfield  {author} {\bibinfo {author} {\bibfnamefont {P.~E.}\ \bibnamefont
  {Bl{\"o}chl}},\ }\href
  {https://journals.aps.org/prb/abstract/10.1103/PhysRevB.50.17953} {\bibfield
  {journal} {\bibinfo  {journal} {Physical Review B}\ }\textbf {\bibinfo
  {volume} {50}},\ \bibinfo {pages} {17953} (\bibinfo {year}
  {1994})}\BibitemShut {NoStop}%
\bibitem [{\citenamefont {Sohier}\ \emph {et~al.}(2017)\citenamefont {Sohier},
  \citenamefont {Calandra},\ and\ \citenamefont {Mauri}}]{sohier2017density}%
  \BibitemOpen
  \bibfield  {author} {\bibinfo {author} {\bibfnamefont {T.}~\bibnamefont
  {Sohier}}, \bibinfo {author} {\bibfnamefont {M.}~\bibnamefont {Calandra}}, \
  and\ \bibinfo {author} {\bibfnamefont {F.}~\bibnamefont {Mauri}},\ }\href
  {https://journals.aps.org/prb/abstract/10.1103/PhysRevB.96.075448} {\bibfield
   {journal} {\bibinfo  {journal} {Physical Review B}\ }\textbf {\bibinfo
  {volume} {96}},\ \bibinfo {pages} {075448} (\bibinfo {year}
  {2017})}\BibitemShut {NoStop}%
\bibitem [{\citenamefont {Leon}\ \emph {et~al.}(2019)\citenamefont {Leon},
  \citenamefont {Costa}, \citenamefont {Chico},\ and\ \citenamefont
  {Latg{\'e}}}]{leon2019interface}%
  \BibitemOpen
  \bibfield  {author} {\bibinfo {author} {\bibfnamefont {C.}~\bibnamefont
  {Leon}}, \bibinfo {author} {\bibfnamefont {M.}~\bibnamefont {Costa}},
  \bibinfo {author} {\bibfnamefont {L.}~\bibnamefont {Chico}}, \ and\ \bibinfo
  {author} {\bibfnamefont {A.}~\bibnamefont {Latg{\'e}}},\ }\href
  {https://www.nature.com/articles/s41598-019-39763-5]} {\bibfield  {journal}
  {\bibinfo  {journal} {Scientific Reports}\ }\textbf {\bibinfo {volume} {9}},\
  \bibinfo {pages} {1} (\bibinfo {year} {2019})}\BibitemShut {NoStop}%
\bibitem [{\citenamefont {Satawara}\ \emph {et~al.}(2021)\citenamefont
  {Satawara}, \citenamefont {Shaikh}, \citenamefont {Gupta},\ and\
  \citenamefont {Gajjar}}]{satawara2021structural}%
  \BibitemOpen
  \bibfield  {author} {\bibinfo {author} {\bibfnamefont {A.~M.}\ \bibnamefont
  {Satawara}}, \bibinfo {author} {\bibfnamefont {G.~A.}\ \bibnamefont
  {Shaikh}}, \bibinfo {author} {\bibfnamefont {S.~K.}\ \bibnamefont {Gupta}}, \
  and\ \bibinfo {author} {\bibfnamefont {P.}~\bibnamefont {Gajjar}},\ }\href
  {https://www.sciencedirect.com/science/article/pii/S2214785320382122}
  {\bibfield  {journal} {\bibinfo  {journal} {Materials Today: Proceedings}\
  }\textbf {\bibinfo {volume} {47}},\ \bibinfo {pages} {529} (\bibinfo {year}
  {2021})}\BibitemShut {NoStop}%
\bibitem [{\citenamefont {Mir{\'o}}\ \emph {et~al.}(2014)\citenamefont
  {Mir{\'o}}, \citenamefont {Audiffred},\ and\ \citenamefont
  {Heine}}]{miro2014atlas}%
  \BibitemOpen
  \bibfield  {author} {\bibinfo {author} {\bibfnamefont {P.}~\bibnamefont
  {Mir{\'o}}}, \bibinfo {author} {\bibfnamefont {M.}~\bibnamefont {Audiffred}},
  \ and\ \bibinfo {author} {\bibfnamefont {T.}~\bibnamefont {Heine}},\ }\href
  {https://pubs.rsc.org/en/content/articlehtml/2014/cs/c4cs00102h} {\bibfield
  {journal} {\bibinfo  {journal} {Chemical Society Reviews}\ }\textbf {\bibinfo
  {volume} {43}},\ \bibinfo {pages} {6537} (\bibinfo {year}
  {2014})}\BibitemShut {NoStop}%
\bibitem [{\citenamefont {Angizi}\ \emph {et~al.}(2020)\citenamefont {Angizi},
  \citenamefont {Akbar}, \citenamefont {Darestani-Farahani},\ and\
  \citenamefont {Kruse}}]{angizi2020two}%
  \BibitemOpen
  \bibfield  {author} {\bibinfo {author} {\bibfnamefont {S.}~\bibnamefont
  {Angizi}}, \bibinfo {author} {\bibfnamefont {M.~A.}\ \bibnamefont {Akbar}},
  \bibinfo {author} {\bibfnamefont {M.}~\bibnamefont {Darestani-Farahani}}, \
  and\ \bibinfo {author} {\bibfnamefont {P.}~\bibnamefont {Kruse}},\ }\href
  {https://iopscience.iop.org/article/10.1149/2162-8777/abb8ef/meta} {\bibfield
   {journal} {\bibinfo  {journal} {ECS Journal of Solid State Science and
  Technology}\ }\textbf {\bibinfo {volume} {9}},\ \bibinfo {pages} {083004}
  (\bibinfo {year} {2020})}\BibitemShut {NoStop}%
\bibitem [{\citenamefont {Huang}\ \emph {et~al.}(2022)\citenamefont {Huang},
  \citenamefont {Grzeszczyk}, \citenamefont {Vaklinova}, \citenamefont
  {Watanabe}, \citenamefont {Taniguchi}, \citenamefont {Novoselov},\ and\
  \citenamefont {Koperski}}]{huang2022carbon}%
  \BibitemOpen
  \bibfield  {author} {\bibinfo {author} {\bibfnamefont {P.}~\bibnamefont
  {Huang}}, \bibinfo {author} {\bibfnamefont {M.}~\bibnamefont {Grzeszczyk}},
  \bibinfo {author} {\bibfnamefont {K.}~\bibnamefont {Vaklinova}}, \bibinfo
  {author} {\bibfnamefont {K.}~\bibnamefont {Watanabe}}, \bibinfo {author}
  {\bibfnamefont {T.}~\bibnamefont {Taniguchi}}, \bibinfo {author}
  {\bibfnamefont {K.}~\bibnamefont {Novoselov}}, \ and\ \bibinfo {author}
  {\bibfnamefont {M.}~\bibnamefont {Koperski}},\ }\href
  {https://journals.aps.org/prb/abstract/10.1103/PhysRevB.106.014107}
  {\bibfield  {journal} {\bibinfo  {journal} {Physical Review B}\ }\textbf
  {\bibinfo {volume} {106}},\ \bibinfo {pages} {014107} (\bibinfo {year}
  {2022})}\BibitemShut {NoStop}%
\bibitem [{\citenamefont {Zhang}\ \emph {et~al.}(2020)\citenamefont {Zhang},
  \citenamefont {Sun}, \citenamefont {Ruan}, \citenamefont {Zhang},
  \citenamefont {Li}, \citenamefont {Zhang}, \citenamefont {Cheng},
  \citenamefont {Wang},\ and\ \citenamefont {Wang}}]{zhang2020point}%
  \BibitemOpen
  \bibfield  {author} {\bibinfo {author} {\bibfnamefont {J.}~\bibnamefont
  {Zhang}}, \bibinfo {author} {\bibfnamefont {R.}~\bibnamefont {Sun}}, \bibinfo
  {author} {\bibfnamefont {D.}~\bibnamefont {Ruan}}, \bibinfo {author}
  {\bibfnamefont {M.}~\bibnamefont {Zhang}}, \bibinfo {author} {\bibfnamefont
  {Y.}~\bibnamefont {Li}}, \bibinfo {author} {\bibfnamefont {K.}~\bibnamefont
  {Zhang}}, \bibinfo {author} {\bibfnamefont {F.}~\bibnamefont {Cheng}},
  \bibinfo {author} {\bibfnamefont {Z.}~\bibnamefont {Wang}}, \ and\ \bibinfo
  {author} {\bibfnamefont {Z.-M.}\ \bibnamefont {Wang}},\ }\href
  {https://aip.scitation.org/doi/full/10.1063/5.0021093} {\bibfield  {journal}
  {\bibinfo  {journal} {Journal of Applied Physics}\ }\textbf {\bibinfo
  {volume} {128}},\ \bibinfo {pages} {100902} (\bibinfo {year}
  {2020})}\BibitemShut {NoStop}%
\bibitem [{\citenamefont {Kim}\ \emph {et~al.}(2017)\citenamefont {Kim},
  \citenamefont {Kim}, \citenamefont {Song}, \citenamefont {Lee},\ and\
  \citenamefont {Lee}}]{kim2017geometric}%
  \BibitemOpen
  \bibfield  {author} {\bibinfo {author} {\bibfnamefont {D.-H.}\ \bibnamefont
  {Kim}}, \bibinfo {author} {\bibfnamefont {H.-S.}\ \bibnamefont {Kim}},
  \bibinfo {author} {\bibfnamefont {M.~W.}\ \bibnamefont {Song}}, \bibinfo
  {author} {\bibfnamefont {S.}~\bibnamefont {Lee}}, \ and\ \bibinfo {author}
  {\bibfnamefont {S.~Y.}\ \bibnamefont {Lee}},\ }\href
  {https://nanoconvergencejournal.springeropen.com/articles/10.1186/s40580-017-0107-0}
  {\bibfield  {journal} {\bibinfo  {journal} {Nano Convergence}\ }\textbf
  {\bibinfo {volume} {4}},\ \bibinfo {pages} {1} (\bibinfo {year}
  {2017})}\BibitemShut {NoStop}%
\bibitem [{\citenamefont {Liu}\ \emph {et~al.}(2003)\citenamefont {Liu},
  \citenamefont {Feng},\ and\ \citenamefont {Shen}}]{liu2003structural}%
  \BibitemOpen
  \bibfield  {author} {\bibinfo {author} {\bibfnamefont {L.}~\bibnamefont
  {Liu}}, \bibinfo {author} {\bibfnamefont {Y.}~\bibnamefont {Feng}}, \ and\
  \bibinfo {author} {\bibfnamefont {Z.}~\bibnamefont {Shen}},\ }\href
  {https://journals.aps.org/prb/abstract/10.1103/PhysRevB.68.104102} {\bibfield
   {journal} {\bibinfo  {journal} {Physical Review B}\ }\textbf {\bibinfo
  {volume} {68}},\ \bibinfo {pages} {104102} (\bibinfo {year}
  {2003})}\BibitemShut {NoStop}%
\bibitem [{\citenamefont {Walla}(2014)}]{walla2014modern}%
  \BibitemOpen
  \bibfield  {author} {\bibinfo {author} {\bibfnamefont {P.~J.}\ \bibnamefont
  {Walla}},\ }\href@noop {} {\emph {\bibinfo {title} {Modern biophysical
  chemistry: detection and analysis of biomolecules}}}\ (\bibinfo  {publisher}
  {John Wiley \& Sons},\ \bibinfo {year} {2014})\BibitemShut {NoStop}%
\bibitem [{\citenamefont {Gali}\ \emph {et~al.}(2009)\citenamefont {Gali},
  \citenamefont {Janz{\'e}n}, \citenamefont {De{\'a}k}, \citenamefont
  {Kresse},\ and\ \citenamefont {Kaxiras}}]{gali2009theory}%
  \BibitemOpen
  \bibfield  {author} {\bibinfo {author} {\bibfnamefont {A.}~\bibnamefont
  {Gali}}, \bibinfo {author} {\bibfnamefont {E.}~\bibnamefont {Janz{\'e}n}},
  \bibinfo {author} {\bibfnamefont {P.}~\bibnamefont {De{\'a}k}}, \bibinfo
  {author} {\bibfnamefont {G.}~\bibnamefont {Kresse}}, \ and\ \bibinfo {author}
  {\bibfnamefont {E.}~\bibnamefont {Kaxiras}},\ }\href
  {https://journals.aps.org/prl/abstract/10.1103/PhysRevLett.103.186404}
  {\bibfield  {journal} {\bibinfo  {journal} {Physical Review Letters}\
  }\textbf {\bibinfo {volume} {103}},\ \bibinfo {pages} {186404} (\bibinfo
  {year} {2009})}\BibitemShut {NoStop}%
\bibitem [{\citenamefont {di~Meo}\ \emph {et~al.}(2009)\citenamefont {di~Meo},
  \citenamefont {Dal~Corso}, \citenamefont {Giannozzi},\ and\ \citenamefont
  {Cozzini}}]{di2009calculation}%
  \BibitemOpen
  \bibfield  {author} {\bibinfo {author} {\bibfnamefont {R.}~\bibnamefont
  {di~Meo}}, \bibinfo {author} {\bibfnamefont {A.}~\bibnamefont {Dal~Corso}},
  \bibinfo {author} {\bibfnamefont {P.}~\bibnamefont {Giannozzi}}, \ and\
  \bibinfo {author} {\bibfnamefont {S.}~\bibnamefont {Cozzini}},\ }\href
  {http://users.ictp.it/~pub_off/lectures/lns024/10-giannozzi/10-giannozzi.pdf}
  {\bibfield  {journal} {\bibinfo  {journal} {Proc. COST School (Trieste)}\ }
  (\bibinfo {year} {2009})}\BibitemShut {NoStop}%
\bibitem [{\citenamefont {Rocha}\ \emph {et~al.}(2018)\citenamefont {Rocha},
  \citenamefont {Rocha}, \citenamefont {Venezuela}, \citenamefont {Garcia},\
  and\ \citenamefont {Ferreira}}]{rocha2018finite}%
  \BibitemOpen
  \bibfield  {author} {\bibinfo {author} {\bibfnamefont {C.~G.~d.}\
  \bibnamefont {Rocha}}, \bibinfo {author} {\bibfnamefont {A.~R.}\ \bibnamefont
  {Rocha}}, \bibinfo {author} {\bibfnamefont {P.}~\bibnamefont {Venezuela}},
  \bibinfo {author} {\bibfnamefont {J.~H.}\ \bibnamefont {Garcia}}, \ and\
  \bibinfo {author} {\bibfnamefont {M.~S.}\ \bibnamefont {Ferreira}},\ }\href
  {https://www.nature.com/articles/s41598-018-27632-6} {\bibfield  {journal}
  {\bibinfo  {journal} {Scientific Reports}\ }\textbf {\bibinfo {volume} {8}},\
  \bibinfo {pages} {1} (\bibinfo {year} {2018})}\BibitemShut {NoStop}%
\bibitem [{\citenamefont {Castleton}\ and\ \citenamefont
  {Mirbt}(2004)}]{castleton2004finite}%
  \BibitemOpen
  \bibfield  {author} {\bibinfo {author} {\bibfnamefont {C.}~\bibnamefont
  {Castleton}}\ and\ \bibinfo {author} {\bibfnamefont {S.}~\bibnamefont
  {Mirbt}},\ }\href
  {https://journals.aps.org/prb/abstract/10.1103/PhysRevB.70.195202} {\bibfield
   {journal} {\bibinfo  {journal} {Physical Review B}\ }\textbf {\bibinfo
  {volume} {70}},\ \bibinfo {pages} {195202} (\bibinfo {year}
  {2004})}\BibitemShut {NoStop}%
\bibitem [{\citenamefont {Castleton}\ \emph {et~al.}(2006)\citenamefont
  {Castleton}, \citenamefont {H{\"o}glund},\ and\ \citenamefont
  {Mirbt}}]{castleton2006managing}%
  \BibitemOpen
  \bibfield  {author} {\bibinfo {author} {\bibfnamefont {C.~W.}\ \bibnamefont
  {Castleton}}, \bibinfo {author} {\bibfnamefont {A.}~\bibnamefont
  {H{\"o}glund}}, \ and\ \bibinfo {author} {\bibfnamefont {S.}~\bibnamefont
  {Mirbt}},\ }\href
  {https://journals.aps.org/prb/abstract/10.1103/PhysRevB.73.035215} {\bibfield
   {journal} {\bibinfo  {journal} {Physical Review B}\ }\textbf {\bibinfo
  {volume} {73}},\ \bibinfo {pages} {035215} (\bibinfo {year}
  {2006})}\BibitemShut {NoStop}%
\bibitem [{\citenamefont {Freysoldt}\ \emph {et~al.}(2009)\citenamefont
  {Freysoldt}, \citenamefont {Neugebauer},\ and\ \citenamefont {Van~de
  Walle}}]{freysoldt2009fully}%
  \BibitemOpen
  \bibfield  {author} {\bibinfo {author} {\bibfnamefont {C.}~\bibnamefont
  {Freysoldt}}, \bibinfo {author} {\bibfnamefont {J.}~\bibnamefont
  {Neugebauer}}, \ and\ \bibinfo {author} {\bibfnamefont {C.~G.}\ \bibnamefont
  {Van~de Walle}},\ }\href
  {https://journals.aps.org/prl/abstract/10.1103/PhysRevLett.102.016402}
  {\bibfield  {journal} {\bibinfo  {journal} {Physical Review Letters}\
  }\textbf {\bibinfo {volume} {102}},\ \bibinfo {pages} {016402} (\bibinfo
  {year} {2009})}\BibitemShut {NoStop}%
\bibitem [{\citenamefont {Komsa}\ \emph {et~al.}(2012)\citenamefont {Komsa},
  \citenamefont {Rantala},\ and\ \citenamefont
  {Pasquarello}}]{komsa2012finite}%
  \BibitemOpen
  \bibfield  {author} {\bibinfo {author} {\bibfnamefont {H.-P.}\ \bibnamefont
  {Komsa}}, \bibinfo {author} {\bibfnamefont {T.~T.}\ \bibnamefont {Rantala}},
  \ and\ \bibinfo {author} {\bibfnamefont {A.}~\bibnamefont {Pasquarello}},\
  }\href {https://journals.aps.org/prb/abstract/10.1103/PhysRevB.86.045112}
  {\bibfield  {journal} {\bibinfo  {journal} {Physical Review B}\ }\textbf
  {\bibinfo {volume} {86}},\ \bibinfo {pages} {045112} (\bibinfo {year}
  {2012})}\BibitemShut {NoStop}%
\bibitem [{\citenamefont {Kwee}\ \emph {et~al.}(2008)\citenamefont {Kwee},
  \citenamefont {Zhang},\ and\ \citenamefont {Krakauer}}]{kwee2008finite}%
  \BibitemOpen
  \bibfield  {author} {\bibinfo {author} {\bibfnamefont {H.}~\bibnamefont
  {Kwee}}, \bibinfo {author} {\bibfnamefont {S.}~\bibnamefont {Zhang}}, \ and\
  \bibinfo {author} {\bibfnamefont {H.}~\bibnamefont {Krakauer}},\ }\href
  {https://journals.aps.org/prl/abstract/10.1103/PhysRevLett.100.126404}
  {\bibfield  {journal} {\bibinfo  {journal} {Physical Review Letters}\
  }\textbf {\bibinfo {volume} {100}},\ \bibinfo {pages} {126404} (\bibinfo
  {year} {2008})}\BibitemShut {NoStop}%
\bibitem [{\citenamefont {Yoo}\ \emph {et~al.}(2021)\citenamefont {Yoo},
  \citenamefont {Todorova}, \citenamefont {Wickramaratne}, \citenamefont
  {Weston}, \citenamefont {Walle},\ and\ \citenamefont
  {Neugebauer}}]{yoo2021finite}%
  \BibitemOpen
  \bibfield  {author} {\bibinfo {author} {\bibfnamefont {S.-H.}\ \bibnamefont
  {Yoo}}, \bibinfo {author} {\bibfnamefont {M.}~\bibnamefont {Todorova}},
  \bibinfo {author} {\bibfnamefont {D.}~\bibnamefont {Wickramaratne}}, \bibinfo
  {author} {\bibfnamefont {L.}~\bibnamefont {Weston}}, \bibinfo {author}
  {\bibfnamefont {C.~G.}\ \bibnamefont {Walle}}, \ and\ \bibinfo {author}
  {\bibfnamefont {J.}~\bibnamefont {Neugebauer}},\ }\href
  {https://www.nature.com/articles/s41524-021-00529-1} {\bibfield  {journal}
  {\bibinfo  {journal} {npj Computational Materials}\ }\textbf {\bibinfo
  {volume} {7}},\ \bibinfo {pages} {1} (\bibinfo {year} {2021})}\BibitemShut
  {NoStop}%
\bibitem [{\citenamefont {Heyd}\ \emph {et~al.}(2003)\citenamefont {Heyd},
  \citenamefont {Scuseria},\ and\ \citenamefont {Ernzerhof}}]{heyd2003hybrid}%
  \BibitemOpen
  \bibfield  {author} {\bibinfo {author} {\bibfnamefont {J.}~\bibnamefont
  {Heyd}}, \bibinfo {author} {\bibfnamefont {G.~E.}\ \bibnamefont {Scuseria}},
  \ and\ \bibinfo {author} {\bibfnamefont {M.}~\bibnamefont {Ernzerhof}},\
  }\href {https://aip.scitation.org/doi/abs/10.1063/1.1564060} {\bibfield
  {journal} {\bibinfo  {journal} {The Journal of Chemical Physics}\ }\textbf
  {\bibinfo {volume} {118}},\ \bibinfo {pages} {8207} (\bibinfo {year}
  {2003})}\BibitemShut {NoStop}%
\bibitem [{\citenamefont {Kubanek}(2022)}]{kubanek2022review}%
  \BibitemOpen
  \bibfield  {author} {\bibinfo {author} {\bibfnamefont {A.}~\bibnamefont
  {Kubanek}},\ }\href {https://arxiv.org/abs/2201.13184} {\bibfield  {journal}
  {\bibinfo  {journal} {arXiv preprint arXiv:2201.13184}\ } (\bibinfo {year}
  {2022})}\BibitemShut {NoStop}%
\end{thebibliography}%
\renewcommand{\theequation}{SI\arabic{equation}}
\setcounter{table}{0}
\setcounter{equation}{0}
\setcounter{figure}{0}
\makeatletter
\renewcommand{\bibnumfmt}[1]{[SI#1]}
\renewcommand{\citenumfont}[1]{SI#1}
\widetext

\end{document}